\documentclass[twocolumn,tighten]{aastex62}
\hypersetup{linkcolor=red,citecolor=blue,filecolor=cyan,urlcolor=magenta}
\usepackage{amsmath}
\usepackage{natbib}
\usepackage{color}
\usepackage{hyperref}
\usepackage{url}
\usepackage{comment}
\usepackage{graphicx}	% Including figure files

\newcommand{\msun}{\mbox{$M_{\odot}$}}

\def\vector#1{\mbox{\boldmath $#1$}}
\newcommand{\pc}{\ensuremath{\,\mathrm{pc}}}
\newcommand{\kpc}{\ensuremath{\,\mathrm{kpc}}}

\newcommand{\Gyr}{\ensuremath{\,\mathrm{Gyr}}}
\newcommand{\kms}{\ensuremath{\,\mathrm{km\ s}^{-1}}}

\newcommand{\masyr}{\ensuremath{\,\mathrm{mas\ yr}^{-1}}}

\newcommand{\vlos}{v_{\ensuremath{\mathrm{los}}}}

\newcommand{\mualpha}{\mu_{\alpha*}}
\newcommand{\mudelta}{\mu_\delta}
\newcommand{\eq}[1]{\begin{align}#1\end{align}}

\newcommand{\new}[1]{{#1}}

\begin{document}

\title{%
Probability of forming gaps in the GD-1 stream by close encounters of globular clusters 
}

\shorttitle{
Globular clusters punching the GD-1 stream
}
%234567890123456789012345678901234567890123
\shortauthors{Doke \& Hattori}

\author[0000-0002-1055-0124]{Yuka~Doke}
\altaffiliation{These authors contributed equally to this work.}
\affiliation{%
Department of Electrical and Electronic Engineering, University of Tokyo, 7-3-1 Hongo, Bunkyo-ku, Tokyo 113-8656, Japan} 
\email{Email:\ dokeyuka@g.ecc.u-tokyo.ac.jp}

\author[0000-0001-6924-8862]{Kohei~Hattori}
\altaffiliation{These authors contributed equally to this work.}
\affiliation{National Astronomical Observatory of Japan, 2-21-1 Osawa, Mitaka, Tokyo 181-8588, Japan}
\affiliation{Institute of Statistical Mathematics, 10-3 Midoricho, Tachikawa, Tokyo 190-8562, Japan}
\affiliation{Department of Astronomy, University of Michigan,
1085 S.\ University Avenue, Ann Arbor, MI 48109, USA}
\email{Email:\ khattori@ism.ac.jp}
\correspondingauthor{Yuka Doke, Kohei Hattori}

\begin{abstract}

One of the most intriguing properties of the GD-1 stellar stream is the existence of three gaps. 
If these gaps were formed by close encounters with dark matter subhalos, the GD-1 stream opens an exciting window through which we can see the size, mass, and velocity distributions of the dark matter subhalos in the Milky Way. 
However, in order to use the GD-1 stream as a probe of the dark matter substructure, we need to disprove that these gaps are not due to the perturbations from baryonic components of the Milky Way. 
Here we ran a large number of test particle simulations 
to investigate the probability that each of the known globular clusters (GCs) can form a GD-1-like gap, by using the kinematical data of the GD-1 stream and GCs from Gaia EDR3 and by fully taking into account the observational uncertainty. 
We found that the probability that all of the three gaps were formed by GCs is as low as 
\new{$1.7\times10^{-5}$}
and the expected number of gaps formed by GCs is only $0.057$ in our fiducial model. 
Our result highly disfavors a scenario in which GCs form the gaps. 
Given that other baryonic perturbers (e.g., giant molecular clouds) are even less likely to form a gap in the retrograde-moving GD-1 stream, 
we conclude that at least one of the gaps in the GD-1 stream was formed by dark matter subhalos if the gaps were formed by flyby perturbations. 
%[239 words]

\end{abstract}
 
%=======================
\section{Introduction}
\label{sec:introduction}

Recent large surveys have revealed dozens of stellar streams in the halo of the Milky Way (MW), 
which are one-dimensional substructures of halo stars 
resulting from disruption of merging stellar systems, such as dwarf galaxies or globular clusters (GCs) \citep{Malhan2018a}. 
These stellar streams not only 
confirm the prediction of the standard $\Lambda$CDM cosmology 
that the galaxies form through hierarchical mergers, 
but also 
allow measurement of the large- and small-scale mass distribution in the Milky Way \citep{Koposov2010,Malhan2019,Bonaca2019,BanikBovy2021}.

The GD-1 stream \citep{Grillmair2006} is one of the most studied stellar streams. 
This stream is characterized by its thinness ($\sim 70 \pc$) 
and its length (more than several tens of degrees) \citep{Carlberg2013}. 
The integrated light from  
\new{the visible part of} 
the GD-1 stream suggest that 
the progenitor system of this stream initially had a mass of $(1.58 \pm 0.07) \times 10^4 M_\odot$ \citep{deBoer2020}, 
\new{
although this mass estimate might be a lower limit if the stream is much more extended than currently recognized.}\footnote{
\new{
Depending on the accretion history of the Milky Way, 
the GD-1 stream may have been affected by long-term gravitational perturbations \citep{Carlberg2020ApJ...889..107C}. 
In such a case, the visible part of the GD-1 stream 
might be a segment of a longer stream; 
and thus the GD-1's progenitor mass might be larger than the estimate by \cite{deBoer2020}. 
}}
These structural properties suggest that its progenitor system 
(which has been completely disrupted) 
was a GC-like system which began disruption at least $\sim 3 \Gyr$ ago \citep{Bowden2015MNRAS.449.1391B,Erkal2016}.

Recently, there have been discoveries of multiple under-dense regions or `gaps' in the GD-1 stream 
\citep{Carlberg2013, deBoer2018, deBoer2020, PriceWhelan2018, Bonaca2019}. 
Although there is some argument that these gaps might be due to the formation process of the stream (such as the non-uniform stripping rate; \new{
epicyclic overdensities,  \citealt{Kupper2010MNRAS.401..105K,Kupper2015ApJ...803...80K}; 
} or the disruption of the progenitor star cluster within a host dwarf galaxy,  \citealt{Malhan2019ApJ...881..106M,Malhan2021MNRAS.501..179M,Qian2022MNRAS.511.2339Q}), 
currently the most widely accepted hypothesis to explain these gaps 
is that these gaps were generated by perturbations from massive compact objects, including the dark matter subhaloes in the MW \citep{Carlberg2009,Carlberg2016,Yoon2011ApJ...731...58Y,Erkal2015a,Erkal2015b,Erkal2016}. 
In general, when a stream experiences a close encounter with a dark matter subhalo with a mass of $\sim 10^5$-$10^7 M_\odot$, the impulsive force from the subhalo results in a differential velocity kick along the stream and forms a gap in the stream. 
Since the gap forming mechanism is well understood, the morphology of the gap and the velocity structure near the gap enable us to investigate the mass and size of the perturber, relative velocity of the perturber with respect to the stream, and the time of the encounter \citep{Erkal2015a}. 
Also, the number of gaps in a stream can put a constraint on the abundance of the dark matter subhalos if all of the gaps are generated by the dark matter subhalos \citep{Erkal2015b}.

To use these gaps as a probe of the dark matter, we need to disprove that these gaps are not generated by baryonic effects, 
such as the perturbation from 
the Galactic bar \citep{Hattori2016,PriceWhelan2016MNRAS.455.1079P}, 
spiral arms of the MW \citep{BanikBovy2021}, 
giant molecular clouds in the stellar disk \citep{Amorisco2016}, 
dwarf galaxies \citep{Bonaca2019}, 
and 
GCs \citep{Bonaca2019,BanikBovy2021}. 
In disproving these effects, 
it is informative to note that the gaps are more difficult to generate 
if a perturber moves at a larger velocity relative to the stream;  
A higher-speed encounter results in a shorter interaction time with the stream and, therefore, a smaller effect. 
Given that the GD-1 stream is orbiting around the MW in a retrograde fashion, 
the perturbations from the bar, spiral arms, and giant molecular clouds -- all of which show a prograde motion -- have negligible effects in generating gaps in the GD-1 stream.\footnote{ 
Another well-studied stream, the Palomar 5 stream, also has some gaps \citep{Erkal2017MNRAS.470...60E}, but it has a prograde orbit.
}
Since the dwarf galaxies are far away from the GD-1 stream at the current epoch and in the past, 
dwarf galaxies also have negligible effect in generating gaps in the GD-1 stream \citep{Bonaca2019}. 
In this regard, 
it is crucial to investigate whether GCs can explain the gaps in the GD-1 stream. 
Previously, \cite{Bonaca2019} tried to pursue this strategy by using the catalog of $\sim 150$ GCs equipped with the astrometric data from Gaia DR2 \citep{Gaia2016A&A...595A...1G, Gaia2018A&A...616A...1G}. 
They integrated the orbit backward in time for 1 Gyr to conclude 
that the impacts from these GCs are negligible in the last 1 Gyr. 
In this paper, we updated their results by using a larger number of GCs (158 GCs) from \cite{Vasiliev2021} equipped with the astrometric data from Gaia EDR3 \citep{Gaia2021A&A...649A...1G}, 
and, more importantly, by adopting a long enough integration time (6 Gyr in our fiducial model). 
We note that \cite{BanikBovy2021} investigated how the baryonic perturbers (including GCs, giant molecular clouds, and spiral arms) affect the density power spectrum along the GD-1 stream by running simulations of the GD-1 stream in the last 3-7 Gyr. 
Our work is complementary to \cite{BanikBovy2021}, because we focus on the {\it probability} that all of the three gaps in the GD-1 stream were formed by GCs.

This paper is organized as follows. 
In Section 2, we present the observed data of the GD-1 stream and the GCs. 
In Section 3, we describe how we set up our test-particle simulations. 
In Section 4, we show the results of our analysis, details of the gap-forming GCs, and the show-case stream models. 
In Section 5, we present some discussion of our results, including the estimated probability that all three gaps in the GD-1 stream were formed by GCs.
In Section 6, we summarize our paper.

%=======================
\section{Data}

Here we describe the data of the GD-1 stream and GCs. 
The coordinate system is defined in Appendix \ref{appendix:coordinate}. 

\subsection{Globular clusters}

Using the astrometric data from Gaia EDR3 \citep{Gaia2021A&A...649A...1G}, 
\cite{Vasiliev2021} derived the position and velocity of 170 GCs in the Milky Way. 
We selected 158 GCs with the full 6D position-velocity data by omitting 12 GCs with incomplete data. 
The data include Right Ascension and Declination $(\alpha, \delta)$, distance $d$, line-of-sight velocity $\vlos$, and proper motion $(\mualpha, \mudelta)$, and their associated uncertainties including the correlation between the two proper motion components. 
We note that the final sample of GCs contains some of the newly discovered GCs for which previous studies (e.g., \citealt{Bonaca2019}) have not checked whether they have experienced a close encounter with the GD-1 stream.

\subsection{Candidate stars of the GD-1 stream}\label{sec:Data_Malhan}

By using STREAMFINDER algorithm \citep{Malhan2018a, Malhan2018b, Malhan2018c}, 
\cite{Malhan2019} compiled a catalog of 97 candidate stars of the GD-1 stream 
(see Table 2 of \citealt{Malhan2019}) 
for which line-of-sight velocity $\vlos$ is taken from 
spectroscopic surveys such as SEGUE \citep{Yanny2009} and LAMOST \citep{Zhao2012}. 
Since their original catalog is based on Gaia DR2 \citep{Gaia2018A&A...616A...1G}, 
we crossmatched these 97 stars with Gaia EDR3 \citep{Gaia2021A&A...649A...1G} 
to obtain more accurate astrometric data. 
Judging from $\vlos$, 
some of their sample stars are probably not a member of the GD-1 stream 
(see Fig. 4d of \citealt{Malhan2019}). 
However, we did not make any manual selection to discard these outliers, 
because such a hard cut might bias our inference on the properties of the GD-1 stream. 
Instead, we modeled the position and velocity of 
the GD-1 candidate stars 
with a mixture model of stream stars and background stars 
(cf., \citealt{Hogg2010arXiv}). 

%============ 
% \cite{Malhan2019} reduced the data to 80 stars after the removal of metal-poor stars with [F/H] $< -1.5$, a few contaminants still remained.
%============

\subsection{Density along the GD-1 stream}

By using the photometric data from Pan-STARRS DR1 and astrometric data from Gaia DR2, 
\cite{deBoer2020} derived the global properties of the GD-1 stream, 
such as the density, distance, and proper motions along the stream. 
They found three gaps along the GD-1 stream 
at $\phi_1 = -36, -20$, and $-3 \deg$ in the 
GD-1 stream coordinate system $(\phi_1, \phi_2)$ 
(see  Appendix \ref{appendix:coordinate}; \citealt{Koposov2010}). 
In this paper, 
we interpret that all of the three gaps are due to some sort of perturbations, 
although some authors interpret that the gap at $\phi_1 = -36 \deg$ 
corresponds to the location of the already-disrupted progenitor (e.g., \citealt{BanikBovy2021}). 
The main goal in this paper is to evaluate 
the probability that three gaps are formed 
within $-40 \deg \leq \phi_1 \leq 0 \deg$ 
by close encounters with known GCs. 
We note that we do {\it not} aim to re-create these three gaps {\it rigorously} at the observed locations.

\begin{figure}
\centering
\includegraphics[width=3.2in]{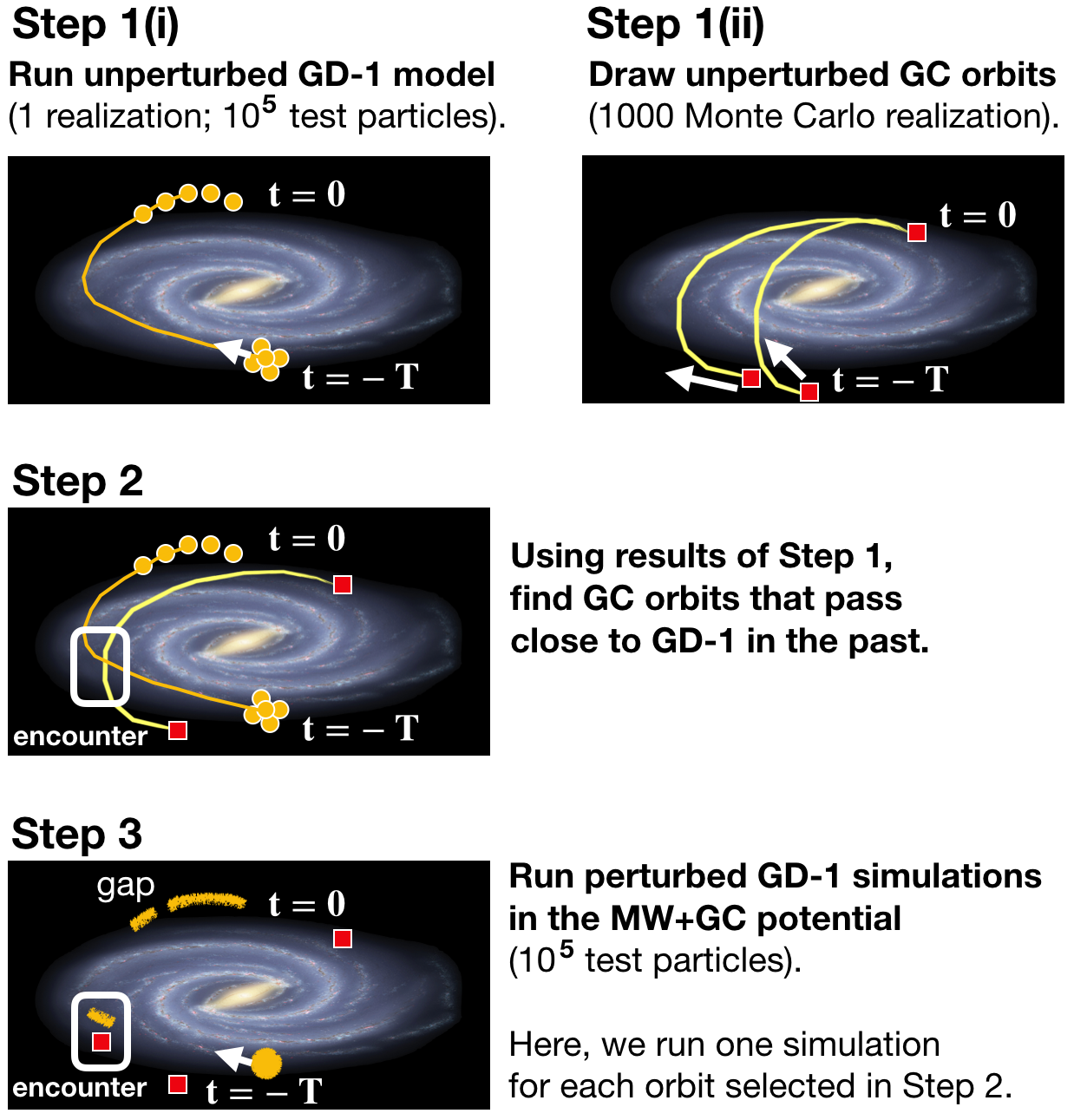}
\caption{Schematic diagram for our simulations in this paper.
}
\label{fig:schematic}
\end{figure}

%=======================
\section{Simulations} \label{sec:simulations}

We analyzed the close encounter with the GD-1 stream and GCs in three steps,  
as described in Fig.~\ref{fig:schematic}. 
%%%%%%%%%
\begin{itemize}
\item Step 1: 
By using a static ({\it unperturbed}) Galactic potential, 
we generated 
(i) an {\it unperturbed} model of the GD-1 stream represented by test particles; 
and 
(ii) $N_\mathrm{MC}=1000$ Monte Carlo orbit models for each GC 
reflecting the observed uncertainties. 
We treat each GC as a test particle moving in the Galactic potential in this step.

\item Step 2: 
By using the {\it unperturbed} models (i) and (ii) in Step 1, 
we select GC orbits that could have encountered the GD-1 stream 
with a small impact parameter and a small relative velocity. 

\item Step 3: 
For each of the selected orbits in Step 2, 
we construct a time-dependent, {\it perturbed} potential, 
consisting of the static MW potential and the time-dependent potential 
from the GC that moves around the MW potential. 
Under this composite MW+GC potential model, 
we ran a test-particle simulation of the GD-1 stream 
{\it perturbed} by a GC due to a close encounter.

\end{itemize}
%%%%%%%%%
Throughout this paper, 
we used the AGAMA package \citep{Vasiliev2019_AGAMA} to run simulations. 
In the following, we describe details of these steps.

\subsection{Model potential of the Milky Way} \label{sec:MWpotential}

In this paper, we used the MW model potential in \cite{McMillan2017}. 
This model is static and axisymmetric, 
and it consists of atomic and molecular gas disks, thin and thick stellar disks, 
bulge, and a dark matter halo. 
In Step 3 of our experiments, we add a time-dependent potential 
caused by the gravitational force from a GC 
to the above-mentioned static MW potential. 
In such a case, we assume 
that the MW potential is rigid and 
that the MW does not wobble, 
because the mass of a GC is much smaller than that of the MW.

\subsection{Step 1 (i): Unperturbed model of the GD-1 stream}

\begin{figure*}
\centering
\includegraphics[width=5.0in]{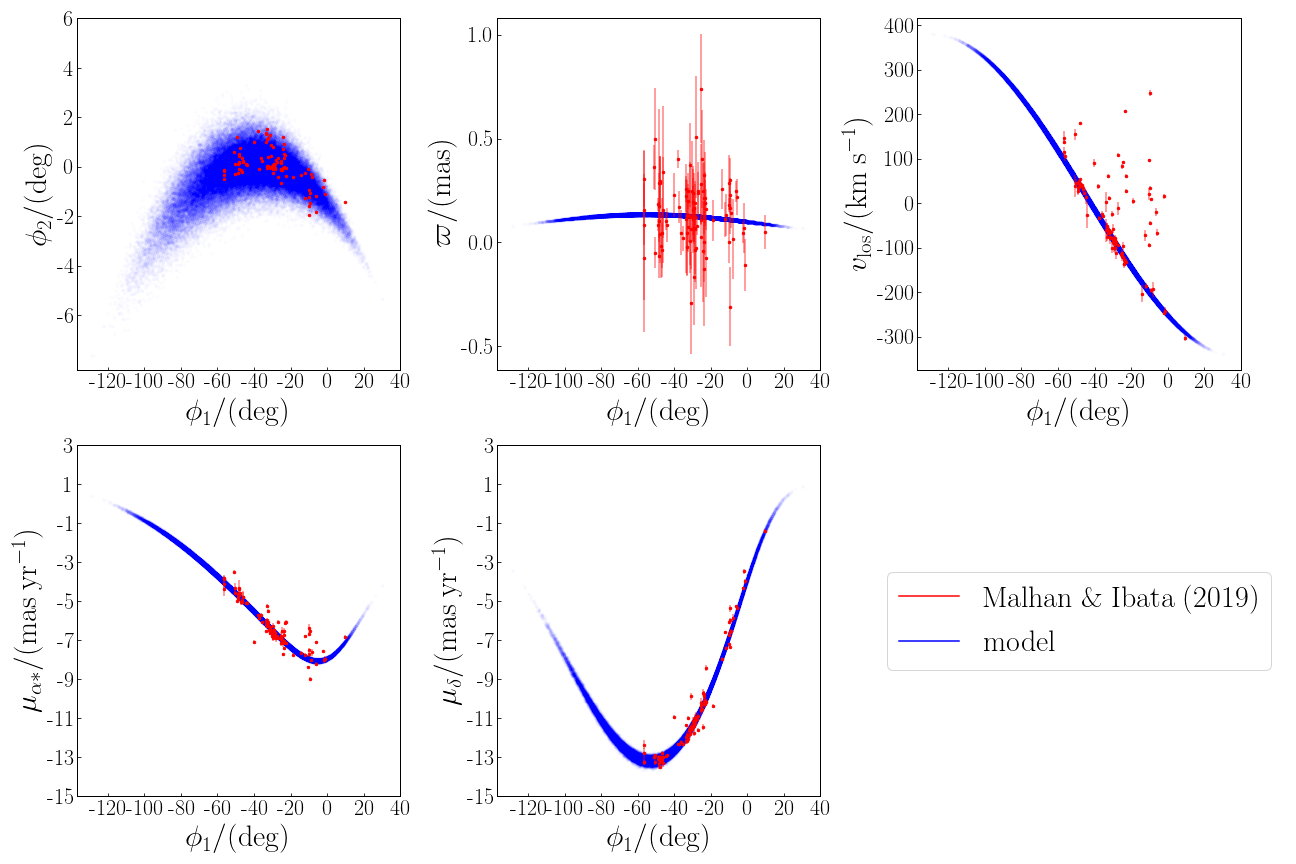}
\caption{
Comparison of the unperturbed GD-1 stream model (blue dots) and 
the observed GD-1 candidate stars in \cite{Malhan2019} (red data points with error bar). 
On each panel, the the horizontal axis is $\phi_1$, 
while the vertical axis is either $(\phi_2, \varpi, \vlos, \mualpha, \mudelta)$. 
As seen in the top right plot showing $(\phi_1, \vlos)$ distribution, 
some of the GD-1 candidate stars 
are outlier (non-member) stars with very different $\vlos$. 
Apart from these outlier stars, 
our unperturbed GD-1 stream model reproduces 
the global properties of the observed GD-1 stream. 
We note that this unperturbed GD-1 stream model 
is used as a reference throughout this paper. 
}
\label{fig:unperturbedGD1}
\end{figure*}

To generate an {\it unperturbed} model of the GD-1 stream, 
we need to assume the orbit of its progenitor system. 
We assume that the GD-1's progenitor 
would be at $\alpha=148.91 \deg$ at the current epoch 
if it had not been completely disrupted. 
This location is chosen following \cite{BanikBovy2021}. 
By assuming that the GD-1 stream approximately delineates the orbit of its progenitor,\footnote{
\new{
In general, the orbit of the progenitor is misaligned with the stream \citep{Sanders2013MNRAS.433.1813S,Sanders2013MNRAS.433.1826S}. 
Although we start from this crude approximation, the final outcome of our simulation more or less reproduces the global properties of the GD-1 stream (see Fig.~\ref{fig:unperturbedGD1}). 
}
} 
we derived the remaining 5D information of the GD-1's progenitor at the current epoch. 
Specifically, 
we fit the 6D phase-space data of the candidates of the GD-1 stream members in \cite{Malhan2019} 
with an orbit by adequately taking into account that some fraction of the stars is non-member stars. 
Our best-fit orbit of the GD-1's progenitor is characterized by  
\eq{
(\alpha^\mathrm{prog}, \delta^\mathrm{prog})|_{t=0}&=(148.91, 36.094) \deg,\\%(148.91, 36.0944389) \deg,\\
d^\mathrm{prog}|_{t=0}&= 7.550 \kpc,\\%7.55009368 \kpc,\\
(\mualpha^\mathrm{prog}, \mudelta^\mathrm{prog})|_{t=0}&=(-5.533,-12.600) \masyr,\\%=(-5.53293288,-12.6006812) \masyr,\\
\vlos^\mathrm{prog}|_{t=0}&=-14.576 \kms. %-14.5757298 \kms ,
}
These quantities are broadly consistent with the result in \cite{BanikBovy2021}. 
\new{
In general, the orbit of the GD-1's progenitor is quite uncertain, not only because the current location $(\alpha^\mathrm{prog}, \delta^\mathrm{prog})|_{t=0}$ 
of the invisible (disrupted) progenitor is arbitrary, but also because the MW potential has evolved over the last several Gyr. 
However, we use our best-fit orbit throughout this paper, 
because our aim is to quantify the encounter rate of the GD-1 stream with GCs and not to reproduce the entire properties of the GD-1 stream. 
}

Given this phase-space information, we integrated the orbit of the GD-1's progenitor backward in time from $t=0$ (current epoch) to $t=-T$. 
We adopt $T=6 \Gyr$ as the fiducial value, 
and thus all the figures in this paper assumes 
this fiducial value. 
We comment on how the choice of $T$ affects our results in Section \ref{sec:dynamical_age}.

Under the model potential we adopted \citep{McMillan2017}, 
the position and velocity of the GD-1's progenitor at $t=-T=-6\Gyr$ are given by\footnote{
See Appendix \ref{appendix:variousT} 
for the corresponding quantites for different choices of $T$. 
} 
%\eq{
%(x, y, z)|_{t=-T} &= (11.939, 12.426, -8.677) \kpc, \\
%(v_x, v_y, v_z)|_{t=-T} &= (-173.465, 53.799, -71.421) \kms. 
%} %-6.0 11.93884 12.426264 -8.67657 -173.4653 53.799473 -71.42131 17.232178
\eq{
&(x^\mathrm{prog}, y^\mathrm{prog}, z^\mathrm{prog})|_{t=-T=-6\Gyr} \nonumber\\
&= (12.079089, 12.387276, -8.601204) \kpc, \\
&(v_x^\mathrm{prog}, v_y^\mathrm{prog}, v_z^\mathrm{prog})|_{t=-T=-6\Gyr} \nonumber\\
&= (-172.22552, 54.99903, -72.57744) \kms. 
} 
%-6.0 12.079089 12.387276 -8.601204 -172.22552 54.99903 -72.57744 17.301704 # Simulation with zero-mass GC004.
The position and velocity of the GD-1's progenitor at $t=-T$ 
was used to create the initial condition of the GD-1 stream particles. 
As a simple prescription to mimic the generation of the GD-1 stream, 
we release $10^5$ test particles at $t=-T$ 
from the same position as the progenitor, 
$(x^\mathrm{prog}, y^\mathrm{prog}, z^\mathrm{prog})|_{t=-T}$, 
with a relative velocity with respect to the progenitor following an isotropic Gaussian distribution:
\eq{
(v_x - v_x^\mathrm{prog})|_{t=-T} &\sim N(0, \sigma_v^2) , \\
(v_y - v_y^\mathrm{prog})|_{t=-T} &\sim N(0, \sigma_v^2) , \\
(v_z - v_z^\mathrm{prog})|_{t=-T} &\sim N(0, \sigma_v^2) . 
} 
Here, $N(0, \sigma^2)$ represents a Gaussian distribution 
with mean $0$ and dispersion $\sigma^2$. 
We regard these $10^5$ particles as the unperturbed GD-1 stream, 
and we integrate the orbit of these $10^5$ particles 
forward in time from $t=-T$ to $t=0$ (current epoch) 
under the unperturbed Galactic potential. 
After some experiments, 
we chose $\sigma_v = (0.5 \kms) (T / 6 \Gyr)^{-1}$ 
so that the length of the GD-1 stream model at the current epoch is 
comparable to the observed extent of the GD-1 stream.  
We recorded the snapshot of these particles 
every 1 Myr and used this information in Step 2.

Fig.~\ref{fig:unperturbedGD1} shows the stellar distribution of 
the unperturbed model at the current epoch. 
We see that the unperturbed model reproduces the observed phase-space 
distribution of the stars except for the apparent outlier stars. 
This unperturbed model is used as the benchmark model 
with which we quantify the strength of perturbation from GCs 
(see Section \ref{sec:defineGD1} and Appendix \ref{appendix:conditionA}). 
We note that randomly chosen 100 stars in the unperturbed model 
is used in Step 2 to find GCs that may have experienced 
a close encounter with the GD-1 stream.

\subsection{Step 1 (ii): Unperturbed orbits of the globular clusters with observational uncertainty}

The GCs in the catalog of \cite{Vasiliev2021} typically have 
$\sim 3$ percent error in distance.
This slight difference makes a noticeable difference 
in evaluating whether a given GC could have experienced a close encounter with the GD-1 stream in the last few Gyr, 
because a slight difference in the initial condition can cause a $\kpc$ scale difference 
in predicting the location of the GC in a few Gyr.

To account for the observational uncertainty, 
we sampled the current position and velocity of each GC for $N_\mathrm{MC}=1000$ times 
from the error distribution. 
Namely, 
we randomly sampled $(d, \mualpha, \mudelta, \vlos)$ for each GC, 
by fully taking into account the correlation between the error in $(\mualpha, \mudelta)$. 
We neglected the small uncertainty in $(\alpha, \delta)$. 
For each GC, we integrated the orbit backward in time from $t=0$ to $t=-T$ 
under the unperturbed Galactic potential. 
We recorded the position and velocity of each GC 
every 1 Myr (and used this information in Step 2). 
For convenience, 
we enumerated each of these Monte Carlo orbits 
by two integers $(j,k)$, 
where $j \; (0 \leq j \leq 169)$ denotes $j$th GC in the catalog of \cite{Vasiliev2021}
and $k \; (0 \leq k \leq 999)$ denotes $k$th Monte Carlo orbit. 
For example, $(j,k)=(4,557)$ 
corresponds to $k=557$th Monte Carlo orbit of NGC1261, 
since NGC1261 is the 4th GC in the catalog.

\subsection{Step 2: Globular clusters with close encounters}

To find GCs that could have experienced a close encounter with the GD-1 stream, 
we used the results in Step 1 (i) and (ii). 
First, from the unperturbed model of the GD-1 stream computed in Step 1 (i), 
we randomly chose 100 particles. 
(We note that we sample 100 particles only once; 
and we use the identical 100 particles throughout Step 2.)
We checked the relative distance and velocity of these 100 GD-1 particles 
with respect to each Monte Carlo orbit of a given GC (computed in Step 1 (ii)) 
as a function of time. 
For each Monte Carlo orbit $(j,k)$ of the GC, 
we searched for a close encounter with the GD-1 stream, 
which we defined as a moment when 
at least one particle in the stream is located within $d_\mathrm{min} < 0.5 \kpc$ from the GC 
having a relative velocity smaller than $v_\mathrm{rel} < 300 \kms$.

As a result, we found that 1383 Monte Carlo orbits of 28 GCs
have experienced a close encounter
with the GD-1 stream in the last $T=6 \Gyr$.
We note that, up to this step, 
it was unclear whether each of these GCs could 
form a gap,\footnote{
For example, some GCs are not massive enough to form a gap. 
} 
which was investigated in Step 3.

%%%%%%%%%%% TABLE  %%%%%%%%%%%
%\startlongtable
\begin{deluxetable*}{l ccc l }
\tablecaption{GCs that can form a gap in the GD-1 stream when $T=6\Gyr$ (fiducial model)
\label{table:probability}}
%\rotate %%Kohei
\tablewidth{0pt}
\tabletypesize{\scriptsize}
\tablehead{
\colhead{Name of GC} &
\colhead{$j$ (GC id)} &
\colhead{$P_\text{gap-forming}$ $^\mathrm{(a)}$} &
\colhead{$N_\text{gap-forming}$ $^\mathrm{(b)}$} &
\colhead{List of $k$ for gap-forming orbits$^\mathrm{(c)}$} %\\
%{} & {(GC id)} & {(Probability of forming} & {(Number of} \\
%{} & {} & {a gap in the GD-1 stream)} & {gap-forming orbits)} 
}
\startdata
NGC1261 & 4 & 0.002 & 2 & 557,591  \\ 
NGC1851 & 9 & 0.001 & 1 & 87  \\ 
NGC2298 & 11 & 0.001 & 1 & 692  \\ 
NGC2808 & 15 & 0.004 & 4 & 43,365,{\bf 698},732  \\ 
NGC3201 & 18 & 0.003 & 3 & {\bf 202},296,435  \\ 
NGC5053 & 31 & 0.001 & 1 & 482  \\ 
NGC5272 (M3) & 34 & 0.022 & 22 & 49,137, 164,175,182,202,251,{\bf 261},391,400  \\ 
\phantom{NGC5272 (M3)} & \phantom{34} & \phantom{0.022} & \phantom{22} & 413,427,515,585,742,744,780,834,841,875  \\ 
\phantom{NGC5272 (M3)} & \phantom{34} & \phantom{0.022} & \phantom{22} & 915,939  \\ 
IC4499       & 40 & 0.009 & 9 & 307,{\bf 311},561,586,670,705,844,934,998  \\ 
NGC5904 (M5) & 45 & 0.001 & 1 & 512  \\ 
NGC6101 & 56 & 0.001 & 1 & 629  \\ 
NGC6229 & 63 & 0.001 & 1 & 208  \\ 
FSR1758 & 89 & 0.003 & 3 & 434,{\bf 731},841  \\ 
NGC6584 & 131 & 0.001 & 1 & 433  \\ 
NGC6864 (M75) & 158 & 0.002 & 2 & 261,849  \\ 
NGC6981 (M72) & 160 & 0.001 & 1 & 78  \\ 
NGC7089 (M2)  & 165 & 0.004 & 4 & 116,{\bf 138},254,534  \\ 
\hline 
\enddata
\tablecomments{
(a) Probability of forming a gap in the GD-1 stream. 
We note that $P_\text{gap-forming} = N_\text{gap-forming}/1000$. 
(b) Number of gap-forming orbits in our simulation. 
We note that we tried 1000 Monte Carlo orbits for each GC. 
(c) Bold-face number corresponds to the show-case example in Table \ref{table:model} and Figs.~\ref{fig:show_case_observables_three_rows_1}-\ref{fig:show_case_observables_three_rows_2}. 
}
\end{deluxetable*}

%%%%%%%%%%%%%%%%%%%%%%%%%%%%%
  
\subsection{Step3: The GD-1 stream models perturbed by globular clusters}

For the 1383 Monte Carlo orbits $(j,k)$ found in Step 2, 
we ran more detailed simulations. 
To efficiently run perturbed model of the GD-1 stream, 
we make some simplifying assumptions.

In each simulation specified by $(j,k)$, 
we only consider one perturber, 
namely the $j$th GC with $k$th Monte Carlo orbit. 
In other words, we do not consider a situation 
where multiple perturbers exist. 
We treat the GD-1 stream particles as test particles 
that feel forces from the MW and $j$th GC (with $k$th Monte Carlo orbit). 
We do not take into account the self-gravity of the GD-1 stream. 
We assume that $j$th GC has a Plummer density profile 
with the total mass $M_{\mathrm{GC},j}$ adopted from 
the compilation by Holger Baumgardt.\footnote{
\url{https://people.smp.uq.edu.au/HolgerBaumgardt/globular/parameter.html}
} 
We assume that the mass of GCs does not change as a function of time. 
We assume that the scale radius of the Plummer profile is $10 \pc$. 
We note that changing the scale radius of GCs in the range of 1-20 $\pc$ 
does not significantly affect our results. 
We assume that the GC only feels the force from the MW, 
and therefore 
the orbit of the GC in Step 2 is identical to that in Step 3. 
We also assume that the MW potential is rigid and does not move 
due to the motion of the GC, 
which is a natural assumption if we consider the perturbation 
from the GC only.\footnote{
In reality, however, 
the MW is not an isolated system 
due to the strong perturbation from the Large Magellanic Cloud 
\citep{Besla2010,Erkal2019,GaravitoCamargo2019,Koposov2019}, 
so our assumption is simplistic in this regard. 
}

Under the above-mentioned assumptions, 
the gravitational potential that 
a GD-1 stream particle feels 
at location ${\bf x}$ and at time $t$ is given by 
\eq{
&\Phi_{\mathrm{total},jk}({\bf x}, t) \nonumber \\
&= \Phi_\mathrm{MW}({\bf x}) + \Phi_{\mathrm{Plummer}} (|{\bf x} - {\bf x}_{\mathrm{GC},jk}(t)|; M_{\mathrm{GC},j}) . \label{eq:Phi_t}
}
Here, 
$\Phi_\mathrm{MW}({\bf x})$ is the static MW potential 
(Section \ref{sec:MWpotential}). 
The $k$th orbit of the $j$th GC is denoted as 
${\bf x}_{\mathrm{GC},jk}(t)$. 
The GC's potential is described by a Pummer potential 
\eq{
\Phi_{\mathrm{Plummer}}(r; M_{\mathrm{GC},j}) = - \frac{GM_{\mathrm{GC},j}}{\sqrt{r^2 + (10 \pc)^2}}.
}

We represent the GD-1 stream with $10^5$ test particles, 
which is large enough to statistically robustly detect a gap in the GD-1 stream model. 
The initial conditions of the GD-1 stream particles 
are chosen in the same manner as in Step 1 (i). 
Importantly, we use the same random seed to generate the initial condition of the GD-1 stream model for all the simulations in Step 1 (i) and Step 3. 
We integrate the orbits of these $10^5$ test particles 
from $t=-T$ to $t=0$ under the perturbed, time-dependent potential 
(equation (\ref{eq:Phi_t})).

After running the simulation, 
we judge whether each stream model contains a GD-1-like gap or not by a simple method in Section \ref{sec:defineGD1}. 
As a result, 57 Monte Carlo orbits of 16 GCs 
result in a GD-1-like gap, 
which will be discussed in Section \ref{sec:result}.

\subsection{Definition of a GD-1-like gap in our simulation} \label{sec:defineGD1}

To assess if a given model in our simulation contains a GD-1-like gap, 
we compare the {\it perturbed} model in Step 3 
with the {\it unperturbed} model in Step 1 (i). 
Because we use the same random seed to create these models, 
whenever we detect a notable difference between these models, 
the difference can be attributed to the effect of the GC perturbation.

To make a fair comparison, 
we first compute the linear density of the unperturbed and perturbed models of the GD-1 stream at the current epoch ($t=0$) 
along $\phi_1$-coordinate, 
$\rho_\mathrm{unperturbed}(\phi_1)$ and 
$\rho_\mathrm{perturbed}(\phi_1)$, respectively, 
by using the histogram of $\phi_1$ with a bin size of $\Delta\phi_1 = 2 \deg$. 
We define that a GD-1-like gap is seen in the model at the location $\phi_1$ 
when the perturbed model stream satisfies the following two conditions: 
\begin{itemize}
\item condition (A) $\frac{\rho_\mathrm{perturbed}(\phi_1)}{ \rho_\mathrm{unperturbed}(\phi_1)} < 0.8$; and 
\item condition (B) $-40 \deg \leq \phi_1 \leq 0 \deg$.   
\end{itemize}
The condition (A) is motivated by the fact that 
a clear gap in our simulations typically satisfy (A). 
The condition (B) is motivated by the fact that 
\cite{deBoer2020} identified three gaps in the GD-1 stream 
at $\phi_1 = -36 \deg, -20 \deg$, and $-3 \deg$. 
We illustrate our procedure in Appendix \ref{appendix:conditionA} and Fig.~\ref{fig:ConditionA}.

%=======================
\section{Result} \label{sec:result}

As described in Section \ref{sec:simulations}, 
we checked in total 158,000 Monte Carlo orbits (1000 orbits for each of the 158 GCs). 
As a result, we found that 57 Monte Carlo orbits of 16 GCs 
%(NGC1261, NGC1851, NGC3201, FSR1758, NGC6981 (M72), NGC7089 (M2)) 
resulted in a gap %(or a clear under-dense region) 
in the GD-1 stream model at $-40 \deg \leq \phi_1 \leq 0 \deg$. 
In Sections \ref{sec:GClist}--\ref{sec:GCorbit}, 
we describe some details of these GCs.
In Section \ref{sec:showcase}, 
we present some show-case examples of the gap-forming models.

\subsection{GC candidates that might have formed a gap in the GD-1 stream} \label{sec:GClist}

Among the 158 GCs that we explored, 
we identified 16 GCs (57 Monte Carlo orbits) 
that formed a GD-1-like gap in the simulation. 
These 16 GCs are good candidates of GCs 
that might have formed the observed gaps in the GD-1 stream. 
These 16 GCs are listed in Table~\ref{table:probability}, 
along with the gap-forming probability $P_{\text{gap-forming}} = N_\text{gap-forming-orbit}/1000$, 
where $N_\text{gap-forming-orbit}$ denotes the number of gap-forming Monte Carlo orbits.

As we see from Table~\ref{table:probability}, $P_{\text{gap-forming}}$ is generally low. 
However, there are 6 GCs that can form a GD-1-like gap with $P_{\text{gap-forming}} \geq \frac{3}{1000}$. 
NGC5272 (M3) and IC4499 are especially interesting GCs, 
which have the highest and second highest probability 
(0.022 and 0.009, respectively) 
of forming a GD-1-like gap.

Most of the 57 Monte Carlo orbits listed in Table~\ref{table:probability} 
have only one close encounter with the GD-1 stream, forming a single gap in the GD-1 stream. 
Intriguingly, 
a few Monte Carlo orbits of NGC5272 (M3) have two close encounters with the GD-1 stream, 
forming one visible gap and another mild under-density region. 
%\kh{Should we put some examples?}

\subsection{Gap-forming probability and the GC mass} \label{sec:GCmass}

\begin{figure}
\centering
\includegraphics[width=3.2in]{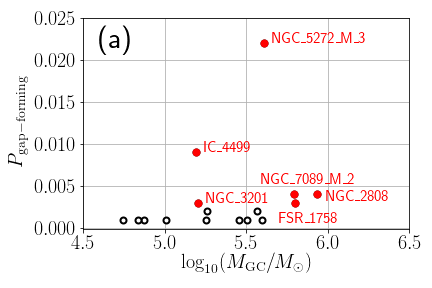}\\
\includegraphics[width=3.2in]{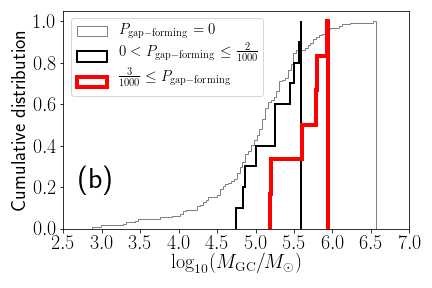}
\caption{
(a) The relationship between the mass and the gap-forming probability 
for the 16 GCs listed in Table \ref{table:probability}. 
The 6 red dots are the GCs with $P_{\text{gap-forming}} \geq \frac{3}{1000}$. 
(b) The normalized cumulative distribution for the mass of GCs with 
different ranges of $P_{\text{gap-forming}}$. 
Those GCs with higher $P_{\text{gap-forming}}$ are typically more massive. 
Also, low-mass GCs ($M_\mathrm{GC} \lesssim 10^{4.5} M_\odot$)  
can hardly form a gap. 
}
\label{fig:mass_prob}
\end{figure}

In Fig.~\ref{fig:mass_prob}(a), 
we show the relationship between the mass and the gap-forming probability 
for the 16 GCs listed in Table \ref{table:probability}. 
The 6 GCs with $P_{\text{gap-forming}} \geq \frac{3}{1000}$ are shown by red filled circle, 
while the 10 GCs with $0<P_{\text{gap-forming}} \leq \frac{2}{1000}$ are shown by black open circle. 
Except for NGC5272 (M3) and IC4499, 
there is a mild trend 
that more massive GCs have a higher gap-forming probability. 
This trend is understandable in the following manner:
Given that 
the maximum impact parameter to form a GD-1-like gap is larger for more massive GCs, more massive GCs have a larger chance of forming a gap. 
In contrast, low-mass GCs need to pass very close to the GD-1 stream to form a gap, and thus lower-mass GCs have a smaller chance of forming a gap.

The 10 GCs with $0<P_{\text{gap-forming}}\leq \frac{2}{1000}$ have only one or two Monte Carlo orbits among 1000 trials 
that result in a GD-1-like gap. 
The mass of these GCs is 
$10^{4.6}M_\odot \lesssim M_\mathrm{GC} \lesssim 10^{5.6}M_\odot$, 
which roughly covers around 20-80 percentiles of the mass of known GCs 
(see gray histogram in Fig.~\ref{fig:mass_prob}(b)). 
Thus, these 10 GCs have a typical mass of the MW GCs. 
Intriguingly, 
the two highest-$P_{\text{gap-forming}}$ GCs,
NGC5272 (M3) and IC4499, 
are not very massive
($10^{5.6}M_\odot$ and $10^{5.2}M_\odot$, respectively). 
Therefore, their high value of $P_{\text{gap-forming}}$ 
is not because they are very massive but because their orbits are favorable to form a gap in the GD-1 stream.

In Fig.~\ref{fig:mass_prob}(b), 
we show the normalized cumulative distribution of the GC mass  
for GCs with different ranges of $P_{\text{gap-forming}}$. 
As we can see from this figure, 
the median GC mass is approximately
$10^{5.0} M_\odot$, $10^{5.25} M_\odot$, and $10^{5.7} M_\odot$, 
for GCs with 
$P_{\text{gap-forming}}=0$, 
$0<P_{\text{gap-forming}} \leq \frac{2}{1000}$, and 
$\frac{3}{1000} \leq P_{\text{gap-forming}}$, respectively. 
This result suggests that 
more massive GCs tend to have a larger probability of forming a gap in the GD-1 stream, supporting the mild trend seen in Fig.~\ref{fig:mass_prob}(a).  
Fig.~\ref{fig:mass_prob}(b) also shows that 
there is no GCs with $M_\mathrm{GC} \lesssim 10^{4.5} M_\odot$ 
that form a gap in the GD-1 stream in our simulation. 
This result suggests that it is extremely difficult ($P_{\text{gap-forming}}<\frac{1}{1000}$) 
to form a gap in the GD-1 stream by these low-mass GCs.

\begin{figure}
\centering
\includegraphics[width=3.2in]{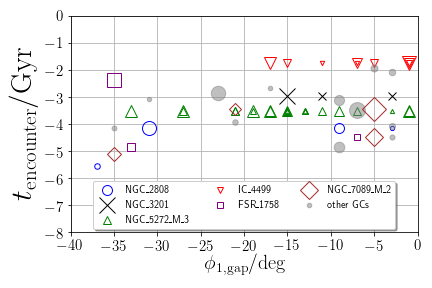}
\caption{
The epoch of the close encounter with a GC ($t_\mathrm{encounter}$) 
and the location of the GD-1-like gap ($\phi_\mathrm{1,gap}$). 
We show all of the 57 simulations listed in Table \ref{table:probability}. 
The 6 GCs with $P_\text{gap-forming} \geq \frac{3}{1000}$ 
are marked by various symbols. 
The other 10 GCs are shown by gray, filled circles. 
The size of the symbol represents the strength of the gap. 
We note that the encounter with NGC5272 (M3) at $t_\mathrm{encounter} \simeq -3.5 \Gyr$ and that with IC4499 at $t_\mathrm{encounter} \simeq -1.7 \Gyr$ can form gaps at $-40 \deg < \phi_1 < 0 \deg $ and $-20 \deg < \phi_1 < 0 \deg$, respectively. 
}
\label{fig:phi1_t}
\end{figure}

\begin{figure}
\centering
\includegraphics[width=3.2in]{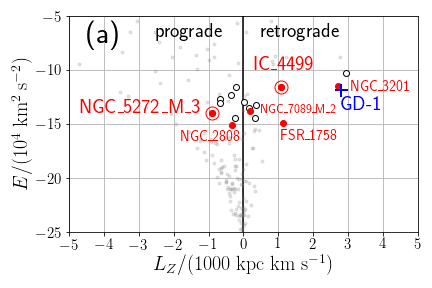}\\
\includegraphics[width=3.2in]{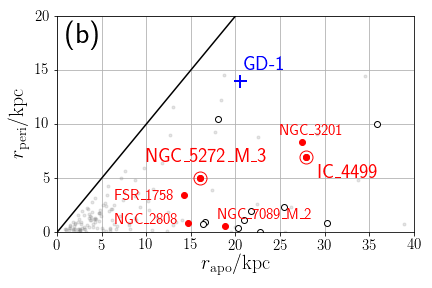}
\caption{
The orbital properties of the GD-1 stream and GCs. 
(a) The azimuthal angular momentum and energy. 
(b) The apocentric and pericentric radii. 
The blue cross ($+$) corresponds to the GD-1 stream. 
The red dots correspond to the 6 GCs with $P_\text{gap-forming} \geq \frac{3}{1000}$; 
among which two GCs (NGC5272 (M3) and IC4499) with the highest-$P_\text{gap-forming}$ are highlighted with red $\odot$. 
The black open circles correspond to the 10 GCs with $0<P_\text{gap-forming} \leq \frac{2}{1000}$. 
The other GCs (namely, those GCs with $P_\text{gap-forming}=0$) are marked with gray dots. 
}
\label{fig:GCorbits}
\end{figure}

\subsection{Epoch of the gap-forming encounters} \label{sec:Encounter_epoch}

Fig.~\ref{fig:phi1_t} shows the relationship between 
the final gap location $\phi_{1,\mathrm{gap}}$ 
and the epoch of the encounter $t_\mathrm{encounter}$ 
for the 6 GCs with $P_\text{gap-forming} \geq \frac{3}{1000}$. 
We see that all of the gap-forming encounters shown here 
happen at $t<-1.5 \Gyr$. 
This result is consistent with the previous work by \cite{Bonaca2019} which claims that none of the GCs seems to have experienced  
a close encounter with the GD-1 stream in the last $1 \Gyr$. 
This result also suggests that 
tracing the past orbits of GCs up to a sufficiently long time ago 
(up to $t=-6 \Gyr$ in our simulation) 
is important to claim 
whether or not GCs can form a visible gap in the GD-1 stream.

Another intriguing result seen in  
Fig.~\ref{fig:phi1_t} is that 
some GCs have their preferable epoch to interact with the GD-1 stream. 
This result is most prominently seen for NGC5272 (M3) and IC4499. 
For example, 
for all of the 22 orbits of NGC5272 (M3) that 
formed a GD-1-like gap, 
the GD-1 stream encounters this GC at $t_\mathrm{encounter} \simeq -3.5 \Gyr$. 
For these 22 orbits, the location of the GD-1 gap in the $\phi_1$ coordinate is distributed at $-40 \deg < \phi_1 < 0 \deg$. 
Thus, an encounter with NGC5272 (M3) at $t_\mathrm{encounter} \simeq -3.5 \Gyr$ can explain any of the observed gaps at $-40 \deg < \phi_1 < 0 \deg$. 
As another example, 
for all of the 9 orbits of IC4499 that 
formed a GD-1-like gap, 
the GD-1 stream encounters this GC at $t_\mathrm{encounter} \simeq -1.7 \Gyr$. 
For these 9 orbits, the location of the GD-1 gap in the $\phi_1$ coordinate is distributed at $-20 \deg < \phi_1 < 0 \deg$, and none of them are distributed at $\phi_1<-20 \deg$. 
Thus, an encounter with IC4499 at $t_\mathrm{encounter} \simeq -1.7 \Gyr$ can explain either of the gaps at $\phi_1=-3 \deg$ or $-20 \deg$ 
but can hardly explain the gap at $\phi_1 = -36 \deg$. 
\new{
To understand these preferable epochs to form a gap, 
we check the orbital phase of the GD-1 stream 
at $t\simeq-3.5 \Gyr$ when it encounters NGC5272 (M3) 
and 
at $t\simeq-1.7\Gyr$ when it encounters IC4499. 
Intriguingly, all of these encounters take place 
when the GD-1 stream is close to its pericenter 
($R\simeq 14 \kpc$). 
The GD-1 stream becomes longest near the pericentric passage, 
and therefore the probability of a close encounter is increased 
due to the enlarged `cross section' of the GD-1 stream. 
As mentioned in Section \ref{sec:GClist}, 
these two GCs (NGC5272 (M3) and IC4499) have the highest probability 
of forming a gap in our fiducial simulations. 
Our finding hints that, in general, GCs that can encounter the GD-1 stream 
near the GD-1's pericenter are a promising perturber to form a gap. 
}

\subsection{Gap-forming probability and the orbit of GCs} \label{sec:GCorbit}

Fig.~\ref{fig:GCorbits} shows the orbital properties of 
the GD-1 stream and GCs investigated in this study. 
In both panels, 
the GD-1 stream is marked by blue cross and 
the 6 GCs with $P_\text{gap-forming} \geq \frac{3}{1000}$ are marked with red dots. 
As we see in Fig.~\ref{fig:GCorbits}(a), 
all of the GCs with $P_\text{gap-forming} \geq \frac{3}{1000}$ have retrograde or mildly prograde orbits 
($L_z > -1000 \kpc \kms$). 
This tendency is understandable because 
the GD-1 stream has a highly retrograde orbit 
with $L_{z,\text{GD-1}} \simeq 2800 \kpc \kms$. 
When a GC with a certain azimuthal angular momentum $L_{z,\text{GC}}$ has a close encounter with the GD-1 stream at a Galactocentric cylindrical radius $R$,
their relative velocity is at least 
$v_\mathrm{rel} \geq |v_{{\phi},\text{GD-1}} - v_{{\phi},\text{GC}}| = |L_{z,\text{GD-1}}-L_{z,\text{GC}}|/R$. 
Given that the radial excursion of the GD-1 stream 
is $14 \kpc \lesssim R \lesssim 20 \kpc$ 
(see Fig.~\ref{fig:GCorbits}(b)), 
those GCs with highly prograde orbits 
$L_{z,\text{GC}} < -2000 \kpc \kms$ 
have at least $v_\mathrm{rel} \gtrsim 200 \kms$. 
Thus, GCs with highly prograde orbits 
are hard form a gap in the GD-1 stream 
unless they are very massive.

We note that 
the GD-1 stream and NGC3201 share similar orbital properties. 
Indeed, \cite{Malhan2022ApJ...926..107M} 
recently claimed that these systems are 
part of the same merging event dubbed `The Arjuna/Sequoia/I’itoi merger' 
(see also \citealt{Bonaca2021ApJ...909L..26B}). 
Although their orbital similarity is intriguing, 
the fact that most of the high-$P_\text{gap-forming}$ GCs 
have very different orbital properties means that 
even if a GC has an orbital property similar to that of the GD-1 stream, 
such a GC is not necessarily a good candidate for forming a gap. 
Rather, as we mentioned in the previous paragraph, 
the relative distance and velocity 
at their closest approach are more important factors to form a gap.

%%%%%%%%%%% TABLE  %%%%%%%%%%%
%\input{table1.tex} 
%\startlongtable
%\begin{rotatetable*}
\begin{deluxetable*}{l l l l l l l }
\tablecaption{Parameters of the GC for six show-case models 
\label{table:model}}
%\rotate %%Kohei
\tablewidth{0pt}
\tabletypesize{\scriptsize}
\tablehead{
\colhead{Model} 
&\colhead{NGC2808\_698}
&\colhead{NGC3201\_202}
&\colhead{NGC5272\_M3\_261}
&\colhead{IC4499\_311}
&\colhead{FSR1758\_731}
&\colhead{NGC7089\_M2\_138}
}
\startdata
Name of the GC & NGC2808& NGC3201& NGC5272 (M3) & IC4499& FSR1758& NGC7089 (M2) \\
$j$ (GC id) & 15 & 18 & 34 & 40 & 89 & 165  \\
$k$ (Monte Carlo id) & 698 & 202 & 261 & 311 & 731 & 138  \\
$M_{\mathrm{GC},j}/M_\odot$ (GC mass) $^\mathrm{(a)}$ & $8.64 \times 10^{5}$ & $1.60 \times 10^{5}$ & $4.06 \times 10^{5}$ & $1.55 \times 10^{5}$ & $6.28 \times 10^{5}$ & $6.20 \times 10^{5}$  \\
\hline 
Current GC observables\\
$\alpha / (\deg)^\mathrm{(b)}$ & $138.013$ & $154.403$ & $205.548$ & $225.077$ & $262.800$ & $323.363$  \\
$\delta / (\deg)^\mathrm{(b)}$ & $-64.863$ & $-46.412$ & $28.377$ & $-82.214$ & $-39.808$ & $-0.823$  \\
$d / (\kpc)$ [model] & $10.1487$ & $4.8280$ & $10.1673$ & $19.0600$ & $13.7330$ & $11.8571$  \\
\phantom{$d / (\kpc)$} [obs.]$^\mathrm{(b)}$ & $10.06 \pm 0.11$ & $4.74 \pm 0.04$ & $10.18 \pm 0.08$ & $18.89 \pm 0.25$ & $11.09 \pm 0.74$ & $11.69 \pm 0.11$  \\
$\mualpha / (\mathrm{mas\;yr^{-1}})$ [model] & $0.9927$ & $8.3694$ & $-0.1624$ & $0.4528$ & $-2.9116$ & $3.3957$  \\
\phantom{$\mualpha / (\mathrm{mas\;yr^{-1}})$} [obs.]$^\mathrm{(b)}$ & $0.995 \pm 0.025$ & $8.351 \pm 0.023$ & $-0.153 \pm 0.024$ & $0.467 \pm 0.025$ & $-2.883 \pm 0.026$ & $3.440 \pm 0.025$  \\
$\mudelta / (\mathrm{mas\;yr^{-1}})$ [model] & $0.2687$ & $-1.9643$ & $-2.6841$ & $-0.4778$ & $2.4720$ & $-2.1981$  \\
\phantom{$\mudelta / (\mathrm{mas\;yr^{-1}})$} [obs.]$^\mathrm{(b)}$ & $0.278 \pm 0.025$ & $-1.972 \pm 0.023$ & $-2.665 \pm 0.023$ & $-0.482 \pm 0.026$ & $2.515 \pm 0.026$ & $-2.165 \pm 0.025$  \\
$\vlos / (\kms)$ [model] & $100.1334$ & $496.3156$ & $-149.9935$ & $35.0794$ & $224.0124$ & $-5.5906$  \\
\phantom{$\vlos / (\kms)$} [obs.]$^\mathrm{(b)}$ & $103.57 \pm 2.00$ & $493.65 \pm 2.00$ & $-147.20 \pm 2.00$ & $38.41 \pm 2.00$ & $227.31 \pm 2.00$ & $-3.78 \pm 2.00$  \\
\hline
Current GC coordinates\\
$x / (\kpc)$ & $-6.0758$ & $-7.5774$ & $-6.7035$ & $2.6558$ & $5.2902$ & $-2.4382$  \\
$y / (\kpc)$ & $-9.7291$ & $-4.7353$ & $1.3378$ & $-14.1939$ & $-2.5651$ & $7.7206$  \\
$z / (\kpc)$ & $-1.9803$ & $0.7253$ & $9.9704$ & $-6.6667$ & $-0.7886$ & $-6.9308$  \\
$v_x / (\kms)$ & $55.5844$ & $253.9416$ & $64.6369$ & $29.4050$ & $250.5375$ & $-72.7037$  \\
$v_y / (\kms)$ & $147.6331$ & $-205.9146$ & $122.1167$ & $245.4252$ & $233.4049$ & $138.3522$  \\
$v_z / (\kms)$ & $30.9628$ & $151.0101$ & $-136.8262$ & $-59.3031$ & $240.6667$ & $-173.9516$  \\
\hline
GC orbital property\\
$L_z / (\kpc \kms)$ & $-356$ & $2762$ & $-905$ & $1069$ & $1877$ & $223$  \\
    & (prograde) & (retrograde) & (prograde) & (retrograde) & (retrograde) & (retrograde) \\
\hline
Output quantities \\
$d_\mathrm{min} / (\kpc)$ & $0.034$ & $0.085$ & $0.060$ & $0.026$ & $0.169$ & $0.030$  \\
$v_\mathrm{rel} / (\kms)$ & $287$ & $200$ & $277$ & $264$ & $290$ & $282$  \\
$t_\mathrm{encounter} / (\Gyr)$ & $-4.154$ & $-2.967$ & $-3.510$ & $-1.739$ & $-2.382$ & $-3.425$  \\
$\phi_{1, \mathrm{gap}}  / (\deg)$ & $-31$ & $-15$ & $-17$ & $-7$ & $-35$ & $-5$  \\
\enddata
\tablecomments{
(a)
Mass of GCs is taken from a compilation by Holger Baumgardt 
(\url{https://people.smp.uq.edu.au/HolgerBaumgardt/globular/parameter.html}). 
(b) 
Observational quantities taken from \cite{Vasiliev2021} 
are shown for a reference. 
}
%\tablenotetext{a}{}
%\startdata
%\enddata
\end{deluxetable*}
%\end{rotatetable*}

%%%%%%%%%%%%%%%%%%%%%%%%%%%%%

\subsection{Details on the show-case models} \label{sec:showcase}

\begin{figure*}
\centering
\includegraphics[width=6.in] {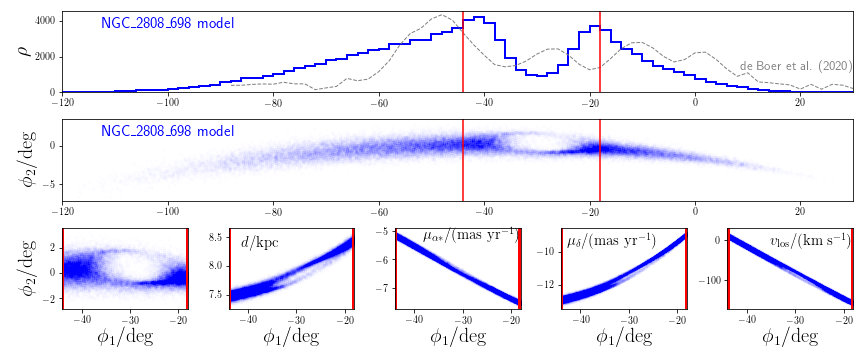} \\
\includegraphics[width=6.in] {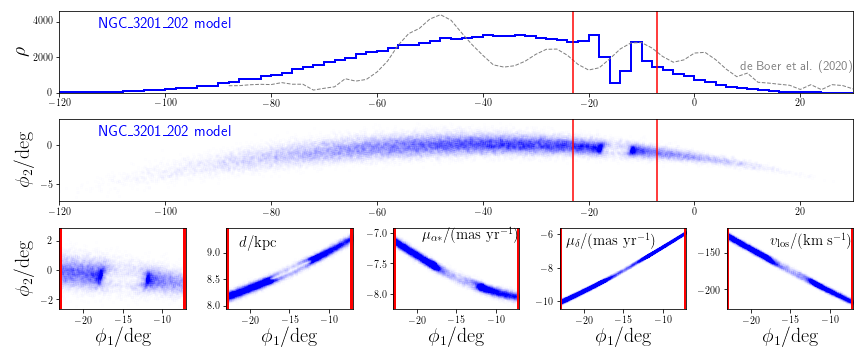} \\
\includegraphics[width=6.in] {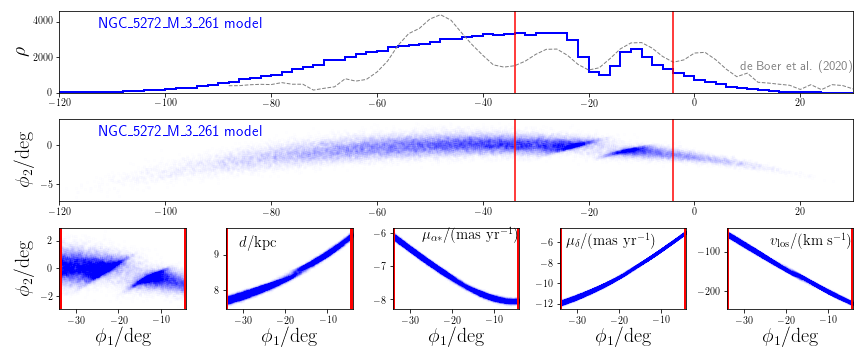} 
\caption{
Morphology of the show-case GD-1 stream models 
NGC2808\_698, NGC3201\_202, and NGC5272\_M3\_261. 
Each model is displayed with three rows, 
and all the horizontal axes are $\phi_1$. 
The upper-most wide panel shows 
the one-dimensional density $\rho(\phi_1)$ in our simulation (blue histogram) 
and in the observed data \citep{deBoer2020} (gray dashed line). 
The middle wide panel shows the full extent of the model stream in $(\phi_1, \phi_2)$ space. 
The lower 5 panels show the model stream near the gap region in $(\phi_1, X)$ space, 
where $X=\phi_2, d, \mualpha, \mudelta$, and $\vlos$, from left to right. 
The red vertical lines surrounding the gap are the same for the upper and lower panels. 
}
\label{fig:show_case_observables_three_rows_1}
\end{figure*}

\begin{figure*}
\centering
\includegraphics[width=6.in] {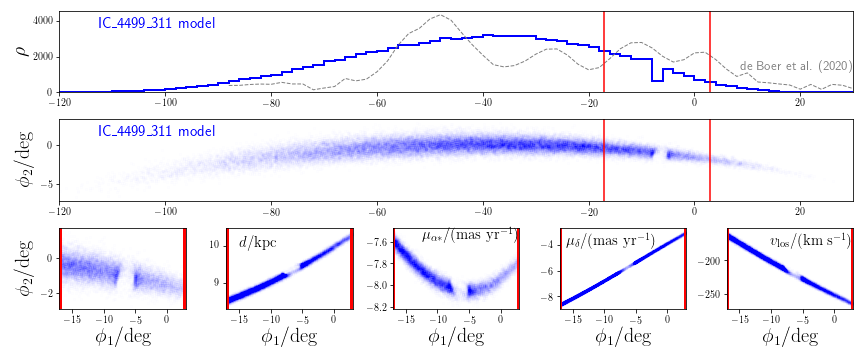} \\
\includegraphics[width=6.in] {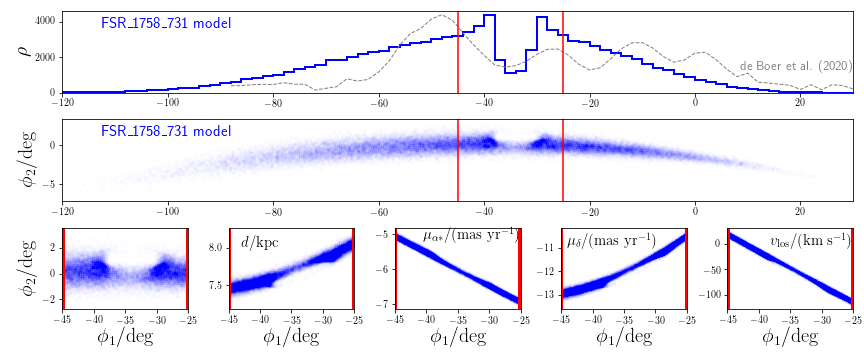} \\
\includegraphics[width=6.in] {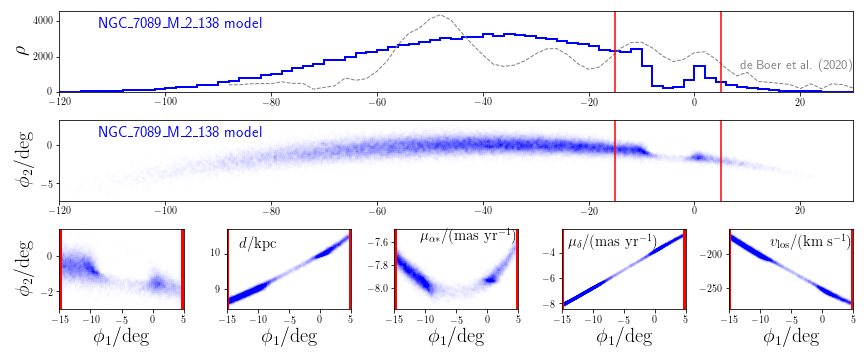} 
\caption{
The same as Fig.~\ref{fig:show_case_observables_three_rows_1}, 
but for the show-case GD-1 stream models 
IC4499\_311, FSR1758\_731, and NGC7089\_M2\_138. 
}
\label{fig:show_case_observables_three_rows_2}
\end{figure*}

For an illustration purpose, for each of the 6 GCs 
with $P_\text{gap-forming} \geq \frac{3}{1000}$, 
we selected one show-case model. 
The details of the selected show-case models 
are summarized in Table~\ref{table:model}. 
The models are named 
{NGC2808\_698}, %$\;$ 
{NGC3201\_202}, $\;$ 
{NGC5272\_M3\_261}, %$\;$ 
{IC4499\_311}, %$\;$ 
{FSR1758\_731}, and
{NGC7089\_M2\_138}. 
Here, 
the last three digits of these names correspond to the value of $k$. 
Figs.~\ref{fig:show_case_observables_three_rows_1} and \ref{fig:show_case_observables_three_rows_2}
show the current-day properties of the GD-1 stream models 
corresponding to these show-case models.

In Figs.~\ref{fig:show_case_observables_three_rows_1} and \ref{fig:show_case_observables_three_rows_2}, 
each show-case model is displayed with three rows. 
The top panel of each show-case model 
shows the one-dimensional density $\rho(\phi_1)$ with an arbitrary unit. 
The blue solid-line histogram shows the histogram of stars in each model. 
The gray dashed line shows the estimated linear density 
(arbitrarily scaled by a constant factor) derived in \cite{deBoer2020}. 
The contrast between the gap and its surrounding over-dense regions in our models is similar to that in the observed GD-1 stream. 
Also, the widths of the gaps in our models are comparable to the observed ones.

The middle and bottom panels in each model in Figs.~\ref{fig:show_case_observables_three_rows_1} and \ref{fig:show_case_observables_three_rows_2} 
shows the morphology of the model in $(\phi_1, X)$ space, 
where $X$ corresponds to various observables. 
The show-case model NGC2808\_698  (Fig.~\ref{fig:show_case_observables_three_rows_1}) 
shows a hole-like structure in $(\phi_1, \phi_2)$ space, 
caused by the perturber (in this case NGC2808) 
that penetrated the stream. 
Interestingly, due to the hole-like structure, 
we see two parallel sequences of the stream 
at $-40 \deg < \phi_1 < -20 \deg$,
which is reminiscent of the observed spur-like feature in the GD-1 stream \citep{PriceWhelan2018, Bonaca2019}. 
A similar hole-like structure in $(\phi_1, \phi_2)$ space 
is also seen in the show-case model NGC3201\_202. 
In this case, a hole-like feature is also seen 
in $(\phi_1, d)$ space, 
indicating that this feature is a three-dimensional structure. 
Given that we see a hole-like structure in multiple models, 
it may be one of the generic features that GCs can form.

Among the 6 GCs, NGC2808, FSR1758, and NGC7089 (M2) are the most massive GCs 
with $M>6\times10^5 M_\odot$. 
Figs.~\ref{fig:show_case_observables_three_rows_1} and \ref{fig:show_case_observables_three_rows_2} 
show that these massive GCs can form a prominent gap, 
even if the relative velocity of the encounter is as large as $\simeq 300 \kms$ (see also Table.~\ref{table:model}). 
In contrast, 
the show-case model IC4499\_311 
results in a clear but narrow gap, 
due to (i) the relatively small mass of IC4499 ($1.55\times 10^5 M_\odot$); 
and (ii) the relatively recent encounter ($\sim 1.7 \Gyr$ ago).

%%%%%%%%%%%%%%%%%%%%%%%%%%%%%%%%%

%=======================
\section{Discussion}

\subsection{Are globular clusters responsible for the GD-1 gaps?} \label{sec:discussion_probability}

The 16 GCs with $P_\text{gap-forming} \geq \frac{1}{1000}$ can possibly form a gap in the GD-1 stream. 
If we assume, as a working hypothesis, that all three gaps in the GD-1 stream were created by the perturbation from these GCs, we can estimate its probability by using $P_\text{gap-forming}$ listed in Table \ref{table:probability}:
% all the probability of ${}_{16} C_3$ combinations. 
%\eq{
%&P_\mathrm{total} \\
%&= \sum_{l\,=\,\mathrm{4,9,...165}}{}\sum_{m\,>\,l}^{}\sum_{n\,>\,m}^{}P_\text{gap-forming}(l)\,P_\text{gap-forming}(m)\,P_\text{gap-forming}(n)\;\;(l, m, n : GC id) \\
%&= 0.0012\%}
\eq{
 &P(\text{GCs formed 3 gaps in the GD-1 stream}) \nonumber \\
=& \sum_{l \in \{4,9, \dots, 165\}} \;\; 
\sum_{m\,>\,l} \;\; 
\sum_{n\,>\,m} 
[
P_\text{gap-forming}(l) \nonumber \\
&
\times P_\text{gap-forming}(m) 
\times P_\text{gap-forming}(n) ] \nonumber \\
=& \new{1.7}\times 10^{-5}. \label{eq:probability}
}
Here, $P_\text{gap-forming}(j)$ corresponds to the 
gap-forming probability of $j$th GC, 
and $\{4,9, \dots, 165\}$ corresponds to the set of $j$ listed in Table~\ref{table:probability}. 
Given this tiny probability, our results suggest that perturbations from GCs are difficult to explain all three gaps in the GD-1 stream. 
Because other baryonic effects 
(e.g., from spiral arms, the Galactic bar, giant molecular clouds, or dwarf galaxies) are even more unlikely to form a gap (see Section \ref{sec:introduction}), 
our results favor a scenario in which at least one of the gaps in the GD-1 stream 
were formed by dark matter subhalos \citep{Carlberg2009,Carlberg2016,Erkal2015a,Erkal2015b,Erkal2016,Bonaca2019,BanikBovy2021}.

Our results can also be used to estimate 
the expected number of gaps formed by GCs in the GD-1 stream, $N_\text{gap}(\text{GCs})$. 
By assuming that a GC can form at most one gap (and no GCs can form multiple gaps), we have
\eq{
N_\text{gap}(\text{GCs}) = \sum_{l \in \{4,9, \dots, 165\}}  P_\text{gap-forming}(l) = 0.057. \label{eq:N_gap}
}
We note that the gaps considered in this paper satisfy condition (A) in Section \ref{sec:defineGD1}, namely  
$\frac{\rho_\mathrm{perturbed}(\phi_1)}{ \rho_\mathrm{unperturbed}(\phi_1)} < 0.8$. 
As a reference, 
\cite{Erkal2016} estimated the 
number of gaps in the GD-1 stream 
formed by dark matter subhalos, $N_\text{gap}(\text{subhalos})$. 
According to their Table~2, 
the expected number of gaps with $\frac{\rho_\mathrm{perturbed}(\phi_1)}{ \rho_\mathrm{unperturbed}(\phi_1)} < 0.75$ 
formed by dark matter subhalos with $(10^5$-$10^9) M_\odot$ 
is $N_\text{gap}(\text{subhalos})=0.6$, 
which is $\sim10$ times larger than our estimate of $N_\text{gap}(\text{GCs})$. 
Although their value of $N_\text{gap}(\text{subhalos})$ is still smaller than $3$ (the observed number of gaps), 
this comparison also favors dark matter subhalos as the cause of the GD-1's gaps.

\subsection{Dynamical age of the GD-1 stream} \label{sec:dynamical_age}

In this paper, we assume that all the stars in the GD-1 stream 
escaped from the progenitor system at $t=-T$. 
The dynamical age of the stream, $T$, 
is set to be $T=6 \Gyr$ in the main analysis of this paper. 
To check how our choice of $T$ affects our result, 
we ran additional simulations with $T= 1, 2, 3$, and $4 \Gyr$ 
(see Appendix \ref{appendix:variousT}). 
(We note that we kept our description in Section \ref{sec:simulations} as general as possible so that the readers can see how the change in $T$ affects the numerical setup of the simulations.) 
For simulations with $T=1 \Gyr$ and $T=2 \Gyr$, 
we found that all the GCs have $P_\text{gap-forming} =0$. 
For simulations with $T=3 \Gyr$, 
we found only two GCs have $P_\text{gap-forming} \geq \frac{1}{1000}$ 
(see Table.~\ref{table:probability_3Gyr_4Gyr}), 
which means that GCs can form at most two gaps. 
For simulations with $T=4 \Gyr$, 
we found 5 GCs with $P_\text{gap-forming} \geq \frac{1}{1000}$ 
(see Table.~\ref{table:probability_3Gyr_4Gyr}). 
By combining the result from our fiducial models with $T=6 \Gyr$, 
we obtain 
\eq{
&P(\text{GCs formed 3 gaps in GD-1}) \nonumber \\
&=
\begin{cases}
0 \phantom{.0\times10^{-8}} \;\; \text{(if $T=1 \Gyr$)}\\
0 \phantom{.0\times10^{-8}} \;\; \text{(if $T=2 \Gyr$)}\\
0 \phantom{.0\times10^{-8}} \;\; \text{(if $T=3 \Gyr$)}\\
6.2\times10^{-8} \;\; \text{(if $T=4 \Gyr$)} \\
\new{1.7}\times10^{-5} \;\; \text{(if $T=6 \Gyr$)}. \label{eq:probability_variousT}
\end{cases}
}
and 
\eq{
N_\text{gap}(\text{GCs}) %\nonumber \\
=
\begin{cases}
0\phantom{.000} \;\; \text{(if $T=1 \Gyr$)}\\
0\phantom{.000} \;\; \text{(if $T=2 \Gyr$)}\\
0.003 \;\; \text{(if $T=3 \Gyr$)}\\
0.010 \;\; \text{(if $T=4 \Gyr$)} \\
0.057 \;\; \text{(if $T=6 \Gyr$)}. \label{eq:N_gap_variousT}
\end{cases}
}
We see that both
the total probability and the expected number of gaps 
become smaller if we adopt a smaller value of $T$. 
We can understand this tendency in two ways. 
First, 
younger streams have fewer opportunities to interact with GCs. 
Second, 
it takes some time for a gap to grow and become visible. 
We note that the previous work by \cite{Bonaca2019} found that 
no GCs experienced a close encounter with the GD-1 stream in the last $1 \Gyr$. 
Their result is consistent with our result with $T=1 \Gyr$. 
Our results suggest that adopting a longer integration time increases 
the chance that GCs can form a gap in the GD-1 stream, 
but it is extremely hard to explain three gaps 
only by the GCs, even if we adopt a long integration time of $T=6\Gyr$.

We note that \cite{BanikBovy2021} investigated the past orbit of the GD-1 stream 
up to $t=-7 \Gyr$. However, they aimed to assess the power spectrum of the density along the GD-1 steam; and not to focus on the individual gaps. 
Thus, our work is complementary to their work.

\subsection{Choice of the model Galactic potential} \label{sec:different_potentials}

\new{
In this paper, we used a model MW potential in \cite{McMillan2017} as the fiducial model. 
In order to check how our results are affected by 
the chosen MW potential, 
we did the same simulations with $T=6 \Gyr$ 
but with potential models in 
\cite{Bovy2015ApJS..216...29B} and \cite{Piffl2014MNRAS.445.3133P}. 
As a result, we found no dramatic changes from our fiducial simulations in the gap-forming probability 
\eq{
&P(\text{GCs formed 3 gaps in GD-1}) \nonumber \\
&=
\begin{cases}
4.8\times10^{-5} \;\; \text{(if $T=6 \Gyr$, \citealt{Bovy2015ApJS..216...29B})} \\
9.7\times10^{-5} \;\; \text{(if $T=6 \Gyr$, \citealt{Piffl2014MNRAS.445.3133P})}  
\label{eq:probability_6Gyr_different_potentials}
\end{cases}
}
and the expected number of gaps
\eq{
&N_\text{gap}(\text{GCs}) \nonumber \\
&=
\begin{cases}
0.076 \;\; \text{(if $T=6 \Gyr$, \citealt{Bovy2015ApJS..216...29B})} \\
0.105 \;\; \text{(if $T=6 \Gyr$, \citealt{Piffl2014MNRAS.445.3133P})}. \label{eq:N_gap_6Gyr_different_potentials}
\end{cases}
}
We note that the list of GCs that form the gaps 
(i.e., the list of GCs in Table \ref{table:probability} in our fiducial model) 
slightly changes if we adopt different MW potentials. 
However, some GCs seem to be more likely to form gaps. 
For example, if we adopt the Galactic potential model in \cite{Bovy2015ApJS..216...29B}, the three important GCs with the highest $P_\text{gap-forming}$ are IC4499, NGC7089 (M2), and NGC3201, all of which appear in Table \ref{table:probability}. 
Also, if we adopt the Galactic potential model in \cite{Piffl2014MNRAS.445.3133P}, the two important GCs with the highest $P_\text{gap-forming}$ are IC4499 and NGC6101, both of which appear in Table \ref{table:probability}. 
Intriguingly, the gap-forming probability of IC4499 is always high $P_\text{gap-forming} \gtrsim 0.01$, 
independent of the adopted potential. 
}

\subsection{Caveats in our analysis} \label{sec:caveats}

As discussed in Section \ref{sec:discussion_probability}, 
we found that it is extremely unlikely that all three gaps in the GD-1 stream are formed by known GCs. 
Here we discuss the limitation of our analysis and possible future directions.

In our simulation, we treated the stars in the GD-1 stream 
as test particles that feel the gravitational force from the MW and a perturbing GC. 
We assume that all the stars were stripped from the 
\new{center of the}
GC-like progenitor system at $t=-T$. 
In reality, 
the stripping process may be continuous, 
\new{
and the stars escape from the progenitor 
from the inner and outer Lagrange points 
\citep{Kupper2010MNRAS.401..105K, Kupper2012MNRAS.420.2700K, Kupper2015ApJ...803...80K, Mastrobuono-Battisti2013MmSAI..84..240M, Sanders2013MNRAS.433.1813S, Sanders2013MNRAS.433.1826S, Bovy2014ApJ...795...95B, Fardal2015MNRAS.452..301F, Guillaume2016MNRAS.460.2711T, Ibata2020ApJ...891..161I}. 
}
Also, we do not take into account the host system of the progenitor system that could affect the morphology of the GD-1 stream 
\citep{Malhan2022ApJ...926..107M,Qian2022MNRAS.511.2339Q}. 
If we were to reproduce the density profile of the GD-1 stream as a function of $\phi_1$, 
the effects mentioned above are important 
and the only way to faithfully take these effects into account is 
to run $N$-body simulations 
\new{
or particle spray simulations (see Appendix \ref{appendix:Lagrange}). 
}
However, because we are interested in the probability that the gaps were formed by the GCs, our approach is good enough for our purpose.

In this paper, we assumed that 
the MW potential is static and axisymmetric. 
These assumptions are simplistic, 
given that the MW is growing in time due to mass accretion \citep{Buist2015}, 
that the MW has a rotating bar \citep{Hattori2016,PriceWhelan2016MNRAS.455.1079P}, 
and that the Large Magellanic Cloud has been perturbing the MW 
\citep{Besla2010,Erkal2019,GaravitoCamargo2019,Koposov2019,Conroy2021Natur.592..534C,Petersen2021NatAs...5..251P,Shipp2021ApJ...923..149S}. 
These effects can alter the values of $P_\text{gap-forming}$ for each GC. 
However, since we did not tune the MW potential to maximize or minimize $P_\text{gap-forming}$ (instead, we just adopted one of the widely-used MW model potentials 
\new{in our main analysis}), 
our estimate of the probability that the GCs formed all three gaps in the GD-1 stream, 
\new{$1.7\times10^{-5}$}
(see equation (\ref{eq:probability})), 
is probably not too far from reality. 
\new{
Indeed, our additional analysis in Section \ref{sec:different_potentials}, in which we varied the potential, supports this view.}
Thus, even though our simulation neglects some important physics, 
we believe our main conclusion is robust: 
the probability that GCs are responsible for all three gaps in the GD-1 stream is extremely low.

%=======================
\section{Conclusion}

In this paper, 
we estimated the probability that Galactic GCs can form a gap in the GD-1 stream 
by using test-particle simulations. 
The summary of this paper is as follows. 

\begin{itemize}
\item 
In our fiducial simulations (in which the GD-1 stream is $T=6 \Gyr$ old), 
16 GCs formed a gap in the GD-1 stream at $-40 \deg < \phi_1 < 0 \deg$ (Table \ref{table:probability}). 
Among them, 6 GCs can form a GD-1-like gap with 
a gap-forming probability 
$P_\text{gap-forming} \geq 0.003$. 
NGC5272 (M3) ($P_\text{gap-forming}=0.022$) and IC4499 ($P_\text{gap-forming}=0.009$) have much higher $P_\text{gap-forming}$ than other GCs. 

\item 
There is a moderate trend that more massive GCs tend to have a larger $P_\text{gap-forming}$ (Fig.~\ref{fig:mass_prob}). 
However, the relative distance and velocity 
\new{
at their closest approach to the GD-1 stream 
}
%However, the relative velocity and distance 
are much more critical factors than the mass of the GCs. 

\item
As shown in Figs.~\ref{fig:show_case_observables_three_rows_1} and ~\ref{fig:show_case_observables_three_rows_2}, our perturbed models can capture some of the morphological properties of the observed GD-1 stream, such as the length, widths, and strength of the gaps. 

\item
The probability that all three gaps in the GD-1 stream 
were formed by the GCs is extremely low. 
In our fiducial model with $T=6 \Gyr$, 
this probability is $P=\new{1.7}\times10^{-5}$ 
(Section \ref{sec:discussion_probability}). 
This probability decreases if we adopt smaller $T$ 
(see equation (\ref{eq:probability_variousT})). 
Assuming $T \leq 3 \Gyr$ results in $P=0$, 
which explains the result of \cite{Bonaca2019} who assumed $T=1 \Gyr$.

\item
The expected number of gaps in the GD-1 stream due to the flyby of GCs 
is $N_\mathrm{gap}(\text{GCs})=0.057$ in our fiducial model (equation (\ref{eq:N_gap})). 
This number is smaller than that due to the flyby of subhalos ($N_\mathrm{gap}(\text{subhalos})=0.6$) by a factor of 10 \citep{Erkal2016}.

\item 
Given (i) that the probability that all three gaps in the GD-1 stream are formed by the GCs is extremely low, 
and (ii) that the retrograde orbit of the GD-1 stream makes other baryonic perturbers (e.g., spiral arms, the Galactic bar, or giant molecular clouds) even less likely to form the gaps, 
at least one of the gaps in the GD-1 stream is probably formed by the 
dark matter subhalos \citep{Carlberg2009,Carlberg2016,Yoon2011ApJ...731...58Y, Erkal2015a,Erkal2015b,Erkal2016,Bonaca2019,BanikBovy2021}. 

\item
To sophisticate our analysis, 
we need to run $N$-body simulations of the encounters of the GD-1 stream and GCs 
by also including the effect from the Large Magellanic Cloud 
\citep{Erkal2019,Koposov2019,Shipp2021ApJ...923..149S}, 
or from the host halo of the GD-1's progenitor 
\citep{Malhan2021MNRAS.501..179M, Qian2022MNRAS.511.2339Q}. 

\end{itemize}

%\input{tab1}
%\phantom{.}
\acknowledgments

The authors thank the referee for thorough reading and constructive comments that improved the original manuscript. 
DY and KH thank NAOJ for financial aid during the 2021 summer student program. 
We thank Junichi Baba for sharing his $N$-body code that gave insights into our work. 
KH thanks lecturers of $N$-body winter school 2021 held by NAOJ for stimulating lectures. 
KH is supported by JSPS KAKENHI Grant Numbers JP21K13965 and JP21H00053.

% [computing]
Numerical computations were in part carried out on GRAPE system at Center for Computational Astrophysics, National Astronomical Observatory of Japan.
This research was supported in part through computational resources and services provided by Advanced Research Computing (ARC), a division of Information and Technology Services (ITS) at the University of Michigan, Ann Arbor.
%
% [observational data]
This work has made use of data from the European Space Agency (ESA) mission
{\it Gaia} (\url{https://www.cosmos.esa.int/gaia}), processed by the {\it Gaia}
Data Processing and Analysis Consortium (DPAC,
\url{https://www.cosmos.esa.int/web/gaia/dpac/consortium}). Funding for the DPAC
has been provided by national institutions, in particular the institutions
participating in the {\it Gaia} Multilateral Agreement.
Guoshoujing Telescope (the Large Sky Area Multi-Object Fiber Spectroscopic Telescope LAMOST) is a National Major Scientific Project built by the Chinese Academy of Sciences. Funding for the project has been provided by the National Development and Reform Commission. LAMOST is operated and managed by the National Astronomical Observatories, Chinese Academy of Sciences.
Funding for the Sloan Digital Sky 
Survey IV has been provided by the 
Alfred P. Sloan Foundation, the U.S. 
Department of Energy Office of 
Science, and the Participating 
Institutions. 
SDSS-IV acknowledges support and 
resources from the Center for High 
Performance Computing  at the 
University of Utah. The SDSS 
website is www.sdss.org.
SDSS-IV is managed by the 
Astrophysical Research Consortium 
for the Participating Institutions 
of the SDSS Collaboration. 
\begin{comment}
including 
the Brazilian Participation Group, 
the Carnegie Institution for Science, 
Carnegie Mellon University, Center for 
Astrophysics | Harvard \& 
Smithsonian, the Chilean Participation 
Group, the French Participation Group, 
Instituto de Astrof\'isica de 
Canarias, The Johns Hopkins 
University, Kavli Institute for the 
Physics and Mathematics of the 
Universe (IPMU) / University of 
Tokyo, the Korean Participation Group, 
Lawrence Berkeley National Laboratory, 
Leibniz Institut f\"ur Astrophysik 
Potsdam (AIP),  Max-Planck-Institut 
f\"ur Astronomie (MPIA Heidelberg), 
Max-Planck-Institut f\"ur 
Astrophysik (MPA Garching), 
Max-Planck-Institut f\"ur 
Extraterrestrische Physik (MPE), 
National Astronomical Observatories of 
China, New Mexico State University, 
New York University, University of 
Notre Dame, Observat\'ario 
Nacional / MCTI, The Ohio State 
University, Pennsylvania State 
University, Shanghai 
Astronomical Observatory, United 
Kingdom Participation Group, 
Universidad Nacional Aut\'onoma 
de M\'exico, University of Arizona, 
University of Colorado Boulder, 
University of Oxford, University of 
Portsmouth, University of Utah, 
University of Virginia, University 
of Washington, University of 
Wisconsin, Vanderbilt University, 
and Yale University.
\end{comment}

\facility{Gaia, LAMOST, SDSS/SEGUE}

\software{
Agama \citep{Vasiliev2019_AGAMA},\;
corner.py \citep{ForemanMackey2016}, 
matplotlib \citep{Hunter2007},
numpy \citep{vanderWalt2011},
scipy \citep{Jones2001}}

\bibliographystyle{aasjournal}
\bibliography{mybibtexfile}
% To add a second bibtex file, separate the name (no suffix) 
% by a comma with no space in the command above.

\appendix

\section{Coordinate system} \label{appendix:coordinate}

We adopt a right-handed Galactocentric Cartesian coordinate system $(x,y,z)$, 
which is the same as in \cite{Hattori2021}. 
The position of the Sun is assumed to be $\vector{x}_\odot = (x_\odot,y_\odot,z_\odot) = (-R_0,0,0)$, with $R_0 = 8.178 \kpc$ \citep{Gravity2019}. The velocity of the Sun with respect to the Galactic rest frame is assumed to be 
$\vector{v}_\odot = (v_{x,\odot},v_{y,\odot},v_{z,\odot}) = (11.10, 247.30, 7.25) \kms$ 
\citep{Reid2004, Schonrich2010}. 
%We also define a Galactocentric spherical coordinate system $(r,\phi,\theta)$ and a Galactocentric cylindrical coordinate system $(R,\phi,z)$, such that $(x,y,z)=(r \cos\theta \cos\phi, r \cos\theta \sin\phi, r \sin\theta) = (R \cos\phi, R \sin\phi, z)$. 
%
%
%Also, for each 3D location with respect to the Sun, we define the line-of-sight unit vector $\vector{e}_\mathrm{los}$. 
Following \cite{Koposov2010}, we define the GD-1 coordinate $(\phi_1, \phi_2)$  such that 
\eq{
\begin{bmatrix} 
\cos\phi_1 \cos\phi_2 \\
\sin\phi_1 \cos\phi_2 \\
\phantom{\sin\phi_1}\sin\phi_2 
\end{bmatrix} 
=
\begin{bmatrix} 
-0.4776303088 & -0.1738432154 & 0.8611897727 \\
0.510844589 & -0.8524449229 & 0.111245042 \\
0.7147776536 & 0.4930681392 & 0.4959603976 
\end{bmatrix} 
\begin{bmatrix} 
\cos\alpha \cos\delta \\
\sin\alpha \cos\delta \\
\phantom{\sin\alpha}\sin\delta 
\end{bmatrix}.
}
The coordinate $\phi_1$ is aligned with the track of the GD-1 stream, 
while the coordinate $\phi_2$ is perpendicular to the stream. 
Due to the historical reason, 
$\phi_1$ decreases along the direction of the GD-1's motion.

\section{Assessment of GD-1-like gaps} \label{appendix:conditionA}

\begin{figure*}
\centering
\includegraphics[width=6.2in] {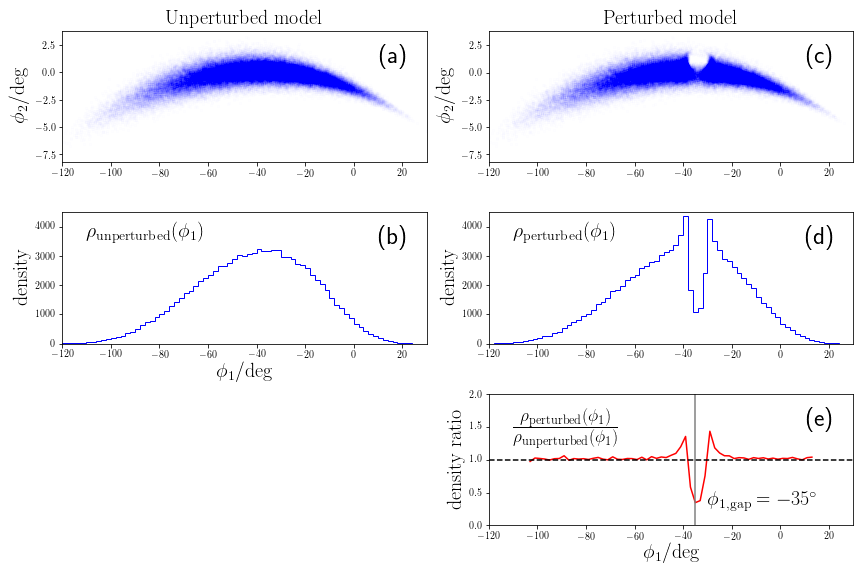} 
\caption{
An illustration of how we compare the unperturbed model with a perturbed model. 
(a) The stellar distribution of the unperturbed model 
in the $(\phi_1, \phi_2)$ space. 
(b) The histogram of the stellar distribution as a function of $\phi_1$. 
This histogram is used as a proxy for the one-dimensional density 
$\rho_\mathrm{unperturbed}(\phi_1)$. 
(c) The same as in panel (a), but for a perturbed model. 
(d) The same as in panel (c), but for the perturbed model. 
This histogram is used as a proxy for the one-dimensional density 
$\rho_\mathrm{perturbed}(\phi_1)$. 
(e) The density ratio $\rho_\mathrm{perturbed}/\rho_\mathrm{unperturbed}$ 
as a function of $\phi_1$. 
The location of the minimum of this ratio (vertical gray line) is defined as the location of the gap, $\phi_\mathrm{1,gap}$. 
If the minimum ratio satisfies   $\rho_\mathrm{perturbed}/\rho_\mathrm{unperturbed}<0.8$ 
and the gap location satisfies $-40 \deg<\phi_\mathrm{1,gap}<0 \deg$, 
we judge that the model has a GD-1-like gap. 
}
\label{fig:ConditionA}
\end{figure*}

As mentioned in Section \ref{sec:defineGD1}, 
we judge whether a given perturbed stream model has a GD-1-like gap 
by comparing the perturbed model with the unperturbed model. 
Fig.~\ref{fig:ConditionA} illustrates how we estimate the strength of the gap 
(under-dense region) in a model stream. 
In this example, we use the show-case model FSR1758\_731. 
%with $(j,k)=(89,731)$. 
This perturbed model has a GD-1-like gap 
at $\phi_\mathrm{1,gap}=-35 \deg$ 
(which satisfies condition (B)). 
The density ratio at $\phi_1=\phi_\mathrm{1,gap}$ is 
$\rho_\mathrm{perturbed}/\rho_\mathrm{unperturbed} = 0.35$, 
which is below our threshold of 0.8 (condition (A)).

\section{Additional simulations with various $T$} \label{appendix:variousT}

The initial conditions of the GD-1's progenitor 
for different $T$ are given as follows:
\eq{
&(x^\mathrm{prog}, y^\mathrm{prog}, z^\mathrm{prog})|_{t=-T}   \nonumber\\
&=
\begin{cases}
(-12.021556, 1.8514469,  9.589011)   \kpc \;\;\text{($T=1\Gyr$)} \\
(-11.69434,  8.299119,  10.201624)   \kpc \;\;\text{($T=2\Gyr$)} \\
(-11.028183, 15.529762,  3.6403887 ) \kpc \;\;\text{($T=3\Gyr$)} \\
(-6.5736866, 17.760225, -7.475801 )  \kpc \;\;\text{($T=4\Gyr$)} \\
\end{cases}
}
\eq{
&(v_x^\mathrm{prog}, v_y^\mathrm{prog}, v_z^\mathrm{prog})|_{t=-T} \nonumber\\
&= 
\begin{cases}
(-108.00474,  -216.09335,  -8.303438) \kms \;\;\text{($T=1\Gyr$)} \\  
(-115.975586, -156.93474,  91.31946)  \kms \;\;\text{($T=2\Gyr$)} \\
(-128.46878,  -72.782036, 129.19073)  \kms \;\;\text{($T=3\Gyr$)} \\
(-156.40662,   -3.031242,  93.21564)  \kms \;\;\text{($T=4\Gyr$)} \\
\end{cases}
}

When $T=1 \Gyr$ and $2\Gyr$ are assumed, 
no GCs form a gap in the GD-1 stream. 
When $T=3 \Gyr$ and $4\Gyr$ are assumed, 
we found that two and five GCs form a gap, respectively, 
as summarized in Table~\ref{table:probability_3Gyr_4Gyr}.

%\startlongtable
\begin{deluxetable*}{l ccc l }
\tablecaption{GCs that can form a gap in the GD-1 stream when $T=3 \Gyr$ and $4\Gyr$ 
\label{table:probability_3Gyr_4Gyr}}
%\rotate %%Kohei
\tablewidth{0pt}
\tabletypesize{\scriptsize}
\tablehead{
\colhead{Name of GC} &
\colhead{$j$ (GC id)} &
\colhead{$P_\text{gap-forming}$ } &
\colhead{$N_\text{gap-forming}$ } &
\colhead{List of $k$ for gap-forming orbits } %\\
%\colhead{$P_\text{gap-forming}$ $^\mathrm{(a)}$} &
%\colhead{$N_\text{gap-forming}$ $^\mathrm{(b)}$} &
%\colhead{List of $k$ for gap-forming orbits$^\mathrm{(c)}$} %\\
%%%{} & {(GC id)} & {(Probability of forming} & {(Number of} \\
%%%{} & {} & {a gap in the GD-1 stream)} & {gap-forming orbits)} 
}
\startdata
$(T=3 \Gyr)$ \\
IC4499 & 40 & 0.002 & 2 & 561,998  \\ 
FSR1758 & 89 & 0.001 & 1 & 731  \\ 
\hline 
$(T=4 \Gyr)$ \\
NGC1851 & 9 & 0.001 & 1 & 87  \\ 
NGC5272 (M3) & 34 & 0.003 & 3 & 366,633,753  \\ 
IC4499 & 40 & 0.004 & 4 & 561,586,844,998  \\ 
FSR1758 & 89 & 0.001 & 1 & 731  \\ 
NGC6981 (M72) & 160 & 0.001 & 1 & 78  \\ 
\hline 
\enddata
%\tablecomments{
%}
\end{deluxetable*}

\clearpage

\section{Comments on the particle spray method} \label{appendix:Lagrange}

\begin{figure*}
\centering
\includegraphics[width=6.2in] {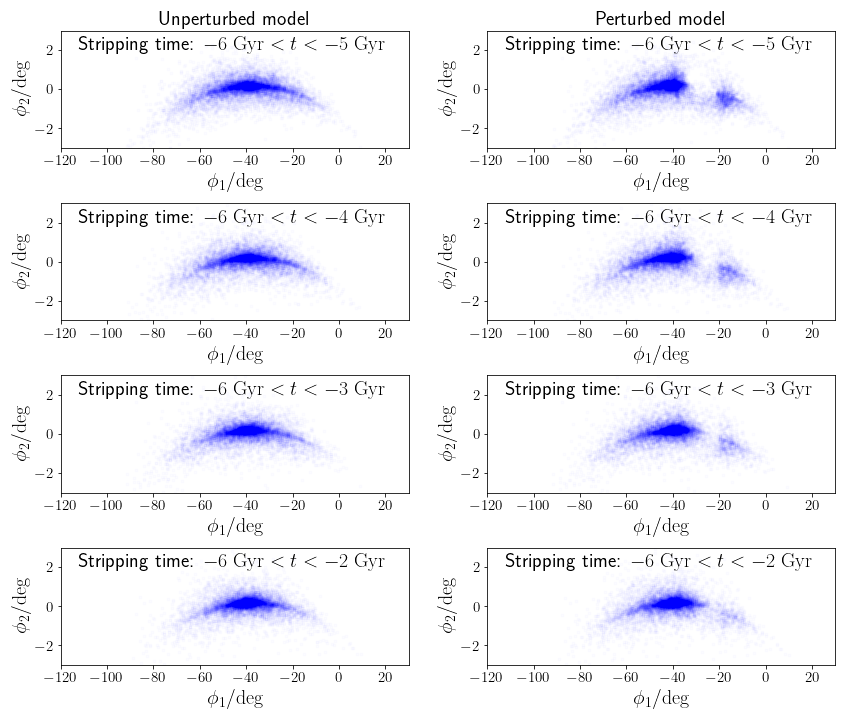} 
\caption{
\new{
Additional simulations of the GD-1 stream model {NGC2808\_698} 
in which we used the particle spray method. 
In each simulation, 
we release the stars from the inner and outer Lagrange points 
with a constant stripping rate at $-6 \Gyr < t < t_\mathrm{end}$. 
From top to bottom, 
we show the results with $t_\mathrm{end} / \Gyr = -5, -4, -3$, and $-2$. The left column corresponds to the simulations with no perturbation. 
The right column corresponds to the simulations with perturbation from NGC2808. 
The gap becomes clearer if the disruption ends at an earlier epoch. 
}
}
\label{fig:Lagrange}
\end{figure*}

\new{
In this paper, we release stream stars at once at $t=-T$, 
from the center of the progenitor. 
In reality, however, stream stars continuously escape from the progenitor 
from the neighborhood of the inner and outer Lagrange points. 
In order to check how our simple prescription affects the results, 
we additionally run eight simulations with the particle spray method, 
mostly following the prescription in 
\cite{Fardal2015MNRAS.452..301F} 
(and partially following \citealt{Kupper2010MNRAS.401..105K}) 
but with a constant stripping rate at $-6 \Gyr < t < t_\mathrm{end}$ 
with $t_\mathrm{end} = (-5,-4,-3,-2) \Gyr$. 
In what follows, 
we focus on 
the show-case model {NGC2808\_698}, 
in which the GD-1 stream and NGC2808 experience a close encounter 
at $t=-4.154 \Gyr$. 
For each $t_\mathrm{end}$, 
we run two simulations 
where the mass of NGC2808 is set to be either 
0 (unperturbed model) or 
$8.64 \times 10^5 \msun$ (perturbed model). 
The result of these simulations is presented in Fig.~\ref{fig:Lagrange}. 
}

\new{
The left-hand panels in Fig.~\ref{fig:Lagrange} 
indicate that the unperturbed streams show a weak tendency 
such that the stream is more concentrated in $\phi_1$-direction 
if the disruption is completed more recently 
(smaller $|t_\mathrm{end}|$). 
This result can be intuitively understood: 
If a progenitor of the stream experience 
a more prolonged disruption history (smaller $|t_\mathrm{end}|$), 
we expect more stars near the progenitor at the current epoch ($t=0$). 
}

\new{
In the right-hand panels in Fig.~\ref{fig:Lagrange}, 
there is a visible gap at $\phi_1 \simeq -30 \deg$ 
for simulations with $t_\mathrm{end}=-5, -4$, and $-3 \Gyr$, 
while 
the gap is almost invisible for the simulation with $t_\mathrm{end}=-2 \Gyr$. 
This result can be understood as follows. 
If $t_\mathrm{end}=-2 \Gyr$, 
a large fraction of stream stars is still bound to the progenitor at $t=-4.154 \Gyr$, 
when the GD-1 stream and NGC2808 experience a close encounter. 
Thus, the number of stars strongly affected by this perturbation is small. 
In addition, some fraction of stars 
that leave the progenitor after $t=-4.154 \Gyr$ 
can reach the gap region and `fill' the gap 
as the stream evolves. 
These two effects make the contrast of the gap weaker 
for simulations with smaller $|t_\mathrm{end}|$. 
}

\new{
The results of the particle spray method indicate 
that the gap is clearer for a larger value of $|t_\mathrm{end}|$, 
or more bursty stripping history. 
In this regard, 
our fiducial simulations -- which corresponds to a bursty stripping or instantaneous disruption -- 
tend to show clearer gaps. 
Therefore, our estimates of the gap-forming probability or the expected number of gaps in the main part of this paper may be an upper limit. 
Given that we find it extremely rare for GCs to form three gaps in the GD-1 stream by using a simplified simulations, 
our additional, more realistic simulations in this Appendix further support our conclusion. 
}

\new{
Lastly, we comment on the difference 
between the particle spray simulations and 
our fiducial simulations. 
The streams generated from the particle spray method 
are more concentrated in $\phi_1$-direction 
than the streams generated in our main analysis of this paper. 
This finding indicates that 
our simplified prescription in the fiducial analysis 
should not be used to interpret the global density profile $\rho(\phi_1)$ of the GD-1 stream. 
(However, we stress that our fiducial models are good enough to 
understand the gap-forming probability.) 
}

\end{document}